\documentclass[iop,english,twocolappendix,numberedappendix,showpacs,superscriptaddress,appendixfloats,tighten,apj,twocolumn]{aastex6}
\nonstopmode
\usepackage{amsmath,amssymb,color,latexsym,natbib,graphicx}
\usepackage[utf8]{inputenc}
\usepackage[T1]{fontenc}
\usepackage{hyperref}
\usepackage{dcolumn,ulem}
\usepackage{subfigure}

\renewcommand{\baselinestretch}{1.2}
\newcommand{\uu}{\textbf{v}}
\newcommand{\vv}{\textbf{v}}
\newcommand{\bb}{\textbf{B}}
\newcommand{\va}{\textbf{v}_{\text{A}}}
\newcommand{\jj}{\textbf{J}}
\newcommand{\jjc}{\textbf{J}_{\text{c}}}
\newcommand{\dv}{\boldsymbol\nabla\cdot\textbf{v}}
\newcommand{\dvp}{\boldsymbol\nabla'\cdot\textbf{v}'}
\newcommand{\dva}{\boldsymbol\nabla\cdot\textbf{v}_{\text{A}}}
\newcommand{\dvap}{\boldsymbol\nabla'\cdot\textbf{v}_{\text{A}}'}
\newcommand{\djc}{\boldsymbol\nabla\cdot\textbf{J}_{\text{c}}}
\newcommand{\djcp}{\boldsymbol\nabla'\cdot\textbf{J}_{\text{c}}'}
\newcommand{\elb}{\boldsymbol{\ell}}
\newcommand{\el}{\boldsymbol{\ell}}
\newcommand{\lang}{\left\langle}
\newcommand{\rang}{\right\rangle}
\newcommand{\mean}{\bar{\delta}}
\newcommand{\nab}{\boldsymbol\nabla}
\newcommand{\eA}{\varepsilon_{A18}}
\newcommand{\eF}{\varepsilon_{F21}}

\newcommand{\dA}{\mathcal{D}_{A18}}
\newcommand{\dF}{\mathcal{D}_{F21}}
\newcommand{\dFm}{\mathcal{D}_{F21}^{loc}}

\newcommand{\fA}{\mathcal{F}_{A18}}
\newcommand{\fF}{\mathcal{F}_{F21}}
\newcommand{\fFm}{\mathcal{F}_{F21}^{loc}}

\newcommand{\dd}{\textbf{d}}

\usepackage{soul}

\begin{document} 

%
%
%
%

\title{An in-depth numerical study of exact laws for compressible Hall magnetohydrodynamic turbulence}

\author{R. Ferrand\altaffilmark{1}, F. Sahraoui\altaffilmark{1}, S. Galtier\altaffilmark{1,2}, N. Andr\'es\altaffilmark{3,4}, P. Mininni\altaffilmark{3} and P. Dmitruk\altaffilmark{3}} 
\email{renaud.ferrand@lpp.polytechnique.fr}
\affil{$^1$ Laboratoire de Physique des Plasmas, CNRS, \'Ecole polytechnique, Universit\'e Paris-Saclay, Sorbonne Universit\'e, Observatoire de Paris-Meudon, F-91128 Palaiseau Cedex, France}
\affil{$^2$ Institut Universitaire de France (IUF)}
\affil{$^3$ Departamento de Física, UBA, Ciudad Universitaria, 1428, Buenos Aires, Argentina}
\affil{$^4$  Instituto de Astronomía y Física del Espacio, CONICET-UBA, Ciudad Universitaria, 1428, Buenos Aires, Argentina}

\date{\today}

\begin{abstract}
Various exact laws governing compressible magnetohydrodynamic (MHD) and Hall-MHD (CHMHD) turbulence have been derived in recent years. Other than their fundamental theoretical interest, these laws are generally used  to estimate the energy dissipation rate from spacecraft observations in order to address diverse problems related, e.g., to heating of the solar wind (SW) and magnetospheric plasmas. Here we use various $1024^3$ direct numerical simulation (DNS) data of free-decay isothermal CHMHD turbulence obtained with the GHOST code (\textit{Geophysical High-Order Suite for Turbulence}) to analyze two of the recently derived exact laws. The simulations reflect different intensities of the initial Mach number and the background magnetic field. The analysis demonstrates the equivalence of the two laws in the inertial range and relates the strength of the Hall effect to the amplitude of the cascade rate at sub-ion scales. When taken in their general form (i.e., not limited to the inertial range) some subtleties regarding the validity of the stationarity assumption or the absence of the forcing in the simulations are discussed. We show that the free-decay nature of the turbulence induces a shift from a large scale forcing towards the presence of a scale-dependent reservoir of energy fueling the cascade or dissipation. The reduced form of the exact laws (valid in the inertial range) ultimately holds even if the stationarity assumption is not fully verified.
\end{abstract}


\section{Introduction}

Understanding the dynamics of turbulent magnetized flows has been a longstanding problem in physics, and especially in astrophysics where turbulence is thought to play a leading role in various physical processes. Examples are the interstellar medium, the SW or planetary magnetospheres, in which turbulence controls structures formation, energy whereabouts and particle heating or acceleration \citep{matthaeus99,bruno13,kritsuk07,arzoumanian11,sahraoui20}. Due to the chaotic nature of turbulence, such media are often studied thanks to the use of specific tools, which rely on statistical methods to uncover trends in the behavior of turbulent flows. A prime example of such tools are exact laws: these equations, which can be obtained through the sole hypothesis of statistical homogeneity (and further refined by introducing time stationarity and infinite Reynolds number) express the rate of energy flowing towards the small/dissipative scales of a system as a function of two-point structure functions, without requiring the use of closure models. Initiated by the work of Kolmogorov and his four-fifths law for hydrodynamic turbulence \citep{kolmogorov}, the quest for exact laws has grown wider ever since with more and more elaborate models being derived. The first steps into studying plasma turbulence were taken by \citet{PP98}, who derived a law for incompressible MHD turbulence. This result paved the way for more precise studies of space plasmas \citep{sorriso07,macbride08,Marino08,Stawarz,Osman,Galtier2012}. More general exact laws have been derived subsequently, by considering the influence of Hall physics in incompressible models \citep{galtier08,BG17,hellinger18,ferrand19}, or the compression of the flow \citep{Carbone09,GB11,banerjee13,BG14,BK17,A2017b,andres18,Banerjee2018,lindborg19,ferrand21,simon2021}.

Thanks to those laws, several studies of astrophysical media have been made possible, either through direct numerical simulations (DNSs) \citep{Mininni09,kritsuk13,verdini15,ferrand20} or \textit{in situ} data analysis \citep{Marino11,coburn15,BanerjeeSW16,hadid17,hadid18,andres19,sorriso19,Bandy2020,andres20}. These allowed testing the efficiency of the exact laws in practical situations in a variety of different systems. Studies specifically designed to test the validity of these exact laws were also conducted, allowing for a more in-depth understanding of how the constituents of these equations relate to each other \citep{Carbone09b,hellinger18,andres18b}. This kind of work is especially important as the models are further refined, and exact laws grow even more complex, such as those derived for compressible Hall MHD, a model taking into account the small scale correction to the plasma dynamics due to the Hall effect. For this last plasma model, where two different exact laws have been derived, such a study has yet to be made. It is thus the aim of this paper to fill this gap by proposing, through the use of an ensemble of high-resolution DNSs for decaying compressible Hall MHD turbulence, a term-by-term analysis of exact laws derived by \citet{andres18} and \citet{ferrand21}. The objective is not only to better understand the exact laws, but to test their more general expressions (valid beyond the inertial range and without time stationarity), which require the bare minimum of hypotheses, and unveil the relations between the energy cascade, the dissipation, and the dynamical variables that form these equations.

The structure of this paper is as follows: first we present the two models from \citet{andres18} and \citet{ferrand21}, which will be referred to as A18 and F21 respectively, and provide compact and general expressions requiring only the assumption of statistical homogeneity to be obtained. We then describe the simulations and the numerical schemes used to compute all the terms forming both exact laws, and then present and discuss the results obtained. Finally, we give a conclusion on our work and on the behavior of the exact laws.

\section{Theoretical models}

Prior to presenting the theory of the two exact laws we introduce some notations and relations tied to the framework of both calculations. We introduce the spatial increment $\el$, connecting two points \textbf{r} and $\textbf{r}'$ in the physical space as 
$\textbf{r}'=\textbf{r}+\el$, and define for any given field $\xi$ : $\xi \equiv \xi(\textbf{r})$ and $\xi' \equiv  \xi(\textbf{r}')$. We also define the notations $\delta\xi = \xi' - \xi$, $\mean \xi = \frac{1}{2}(\xi'+\xi)$ and the differential operator in the direction $\el$ as $\nab_{\el}$. This operator obeys the following relation on ensemble averages $\lang \rang$ : $\langle\nab'\cdot\rangle=-\langle\nab\cdot\rangle=\nab_{\elb} \cdot\langle\rangle$.

\subsection{F21 model}

The isothermal CHMHD exact law derived by \citet{ferrand21} is obtained by considering the following three-dimensional compressible HMHD equations:
\begin{align} 
	\partial_t \rho + \nab \cdot (\rho \uu) =& \, 0 \, , \label{MHD1_F21} \\
        \rho ( \partial_t \uu + \uu\cdot\nab\uu ) =& - \nab P+ \jj \times \bb + {\bf d_{\nu}} +{\bf f} \label{MHD2_F21} \, , \\
        \partial_t \bb =& \nab \times (\uu \times \bb) - \lambda \nab \times (\jjc \times \bb) + {\bf d_{\eta}} \, , \label{MHD3_F21} \\
        \nab \cdot \bb =& \, 0 \, , \label{MHD4_F21}
\end{align}
where $\rho$ is the mass density, $\uu$ the velocity, $P$ the pressure, $\bb$ the magnetic field, $\jj=(\nab \times \bb)/\mu_{0}$ the current density and $\jjc=\jj/\rho$ the normalized current. The Hall effect is introduced in the model through the presence of the term $ \lambda \nab \times (\jjc \times \bb)$ in the induction equation: $\lambda=m_i/q_e$, with $m_{i}$ the ion mass and $q_e$ the magnitude of the electron charge, is the marker of the Hall effect and is connected to the ion inertial length $d_i$ through the relation $d_i=\lambda/\sqrt{\mu_0 \rho}$, with $\mu_0$ the vacuum permeability.
 The dissipation terms are: 
\begin{align} 
{\bf d_{\nu}} =& \, \nu \Delta \uu + \frac{\nu}{3} \nab \theta \, , \\
{\bf d_{\eta}} =& \, \eta \Delta \bb \, , 
\end{align} 
with $\theta = \nab \cdot \uu$ the dilatation, $\nu$ the dynamic viscosity and $\eta$ the magnetic diffusivity. The term $\textbf{f}$ represents a large scale forcing. The isothermal closure writes $P= c_s^2 \rho$ 
with $c_{s}$ the constant speed of sound. These equations are used to derive a dynamical equation for the modified second-order structure function
\begin{align}
    \langle S \rangle \equiv& \left\langle \frac{1}{2} \mean \rho |\delta \uu|^2+ \frac{1}{2\mu_0} |\delta \bb|^2 + \frac{1}{2} \delta \rho \delta e \right\rangle . \label{sfF21}   
\end{align}	
In the simulation code the internal energy follows the polytropic definition $e= \frac{c_{s}^{2}}{\gamma (\gamma-1)}((\rho/\rho_0)^{\gamma-1}-1)$ with a polytropic index $\gamma = 1.01$ close enough to unity so that the isothermal approximation remains valid. Note that under this approximation entropy is assumed to be constant, and thus the internal energy variation results only from the work of the pressure force \citep{simon2021}.

Injecting equations (\ref{MHD1_F21})-(\ref{MHD4_F21}) in the time derivative of (\ref{sfF21}) leads, after a hefty amount of calculations, to the equation:
\begin{widetext}
\begin{align}
\partial_t \langle S \rangle =& \partial_t \langle E^{tot} \rangle - \frac{1}{2} \nab_{\elb} \cdot \left \langle \mean \rho \vert\delta \uu\vert^2 \delta \uu \right \rangle 
+ \frac{1}{4} \langle (\rho \theta' + \rho' \theta) \vert\delta \uu \vert^2 \rangle 
- \langle \delta \rho \, \delta \uu \cdot \mean (\jjc \times \bb) \rangle + {\lambda} \langle \delta (\jj \times \bb) \cdot \delta \jjc  \rangle  \nonumber \\
&- \frac{1}{2\mu_0} \nab_{\elb} \cdot \left\langle \vert \delta \bb \vert^2 \delta \uu - 2 (\delta \uu \cdot \delta \bb) \delta \bb 
- \lambda \vert \delta \bb \vert^2 \delta \jjc + 2 \lambda(\delta \bb \cdot \delta \jjc) \delta \bb \right \rangle \nonumber \\
&- \frac{1}{2} \left\langle \left(1+ \frac{\rho'}{\rho} \right) \uu' \cdot ({\bf d_\nu} + {\bf f}) + \left(1+\frac{\rho}{\rho'}\right) \uu \cdot ({\bf d_\nu'} + {\bf f'}) \right\rangle \nonumber \\
&+ \frac{1}{2} \left\langle \frac{\rho}{\rho'} \uu' \cdot ({\bf d_\nu'} + {\bf f'}) + \frac{\rho'}{\rho} \uu \cdot ({\bf d_\nu} + {\bf f}) \right\rangle - \frac{1}{\mu_0} \langle \bb' \cdot {\bf d_{\eta}} + \bb \cdot {\bf d'_{\eta}} - \bb \cdot {\bf d_\eta} \rangle \, , \label{fullF21}
\end{align}
\end{widetext}
with the total energy $E^{tot} =  \rho u^{2}/2 + B^{2}/(2\mu_{0}) + \rho e$. Equation (\ref{fullF21}), already reported in \citet{ferrand21}, is the most general equation obtainable in this model under the sole assumption of statistical homogeneity. With additional assumptions of time stationarity, forcing limited to the largest scales and infinite (magnetic and kinetic) Reynolds numbers one can then retrieve the exact law:
\begin{widetext}
\begin{align} \nonumber
- 4 \eF =& \nab_{\el} \cdot \left \langle \mean \rho \vert\delta \uu\vert^2 \delta \uu + \frac{1}{\mu_0} \vert \delta \bb \vert^2 \delta \uu - \frac{2}{\mu_0} (\delta \uu \cdot \delta \bb) \delta \bb \right \rangle - \frac{1}{2} \langle (\rho \theta' + \rho' \theta) \vert \delta \uu \vert^2 \rangle + 
2 \langle \delta \rho \, \delta \uu \cdot \mean (\jjc \times \bb) \rangle \\
+& \frac{\lambda}{\mu_0} \nab_{\el} \cdot \left\langle 2 (\delta \bb \cdot \delta \jjc) \delta \bb - \vert \delta \bb \vert^2 \delta \jjc \right \rangle - 2{\lambda} \langle \delta (\jj \times \bb) \cdot \delta \jjc \rangle . \label{F21}
\end{align}
\end{widetext}
Equation (\ref{F21}) (or similar ones obtained for other models) is generally the one that is used in numerical simulations and spacecraft data to infer the cascade rate (left-hand side) from measurable quantities (right-hand side) \citep{banerjee16,hadid17,hadid18,andres18b,andres19}. However, to be obtained, non-trivial approximations had to be made, about time stationarity and on the forcing and dissipative terms \citep{ferrand21}. While these approximation are hard (if not impossible) to test on spacecraft data, they can {\it a priori} be verified in DNSs, which is one of the goals of the present study. The specific point about time stationarity and the absence of driving is of a particular relevance in free-decay simulations as the ones we are using in this study. Therefore, we will consider in the following the more general equation (\ref{fullF21}), written in a more compact form:
\begin{equation} \label{fullF21_comp}
2\eF + \partial_t \lang E^{tot} \rang - \partial_t \lang S \rang + \dF + \dFm + \fF + \fFm = 0
\end{equation}
where we define:
\begin{widetext}
\begin{align}\label{DF21}
    \dF \equiv& -\frac{1}{2}\lang \left(1+ \frac{\rho'}{\rho} \right) \uu' \cdot {\bf d_\nu} + \left(1+\frac{\rho}{\rho'}\right) \uu \cdot {\bf d_\nu'} \rang -  \frac{1}{\mu_0}\lang \bb' \cdot {\bf d_{\eta}} + \bb \cdot {\bf d'_{\eta}} \rang , \\
    \dFm \equiv&~ +\frac{1}{2} \lang \frac{\rho}{\rho'} \uu' \cdot {\bf d_\nu'} + \frac{\rho'}{\rho} \uu \cdot {\bf d_\nu} \rang + \frac{1}{2\mu_0} \lang \bb \cdot {\bf d_\eta} + \bb' \cdot {\bf d'_\eta} \rang , \\
     \fF \equiv& - \frac{1}{2} \left\langle \left(1+ \frac{\rho'}{\rho} \right) \uu' \cdot {\bf f} + \left(1+\frac{\rho}{\rho'}\right) \uu \cdot {\bf f'} \right\rangle , \\
     \fFm \equiv& + \frac{1}{2} \left\langle \frac{\rho}{\rho'} \uu' \cdot {\bf f'} + \frac{\rho'}{\rho} \uu \cdot {\bf f} \right \rangle.
\end{align}
\end{widetext}
Note that in equation (\ref{fullF21_comp}) $\eF$ is introduced by identifying it in equation (\ref{fullF21}) to its expression given by relation (\ref{F21}). It {\it does not} result from using the same assumptions that led to relation (\ref{F21}), and thus equation (\ref{fullF21_comp}) remains very general.
The superscript {\it loc} is used for the forcing and dissipation terms that result from the product of vectors taken locally (i.e., at the same position \textbf{r} or $\textbf{r}'$), assuming that $\rho\sim \rho'$ for the velocity field terms. This contrasts with  the terms $\dF$  and $\fF$ that involve (distant) two-point correlations. Assuming that $\rho\sim \rho'$ in weakly compressible simulations such as those of this study, the terms $\dF$ and $\fF$ are likely to be scale-independent. However, as we will show below they still have a significant impact on the energy balance. The same remark can be made about the term $\partial_t \lang E^{tot} \rang$ in Eq. \ref{fullF21_comp}, which is clearly a local term (unlike $\partial_t \lang S \rang$ since $S$ is a second order structure function).

\subsection{A18 model}
The other exact law derived by \citet{andres18} relies on the same three-dimensional isothermal HMHD equations, yet using the Alfv\'en speed $\textbf{v}_A \equiv \textbf{B}/\sqrt{\mu_0\rho}$ instead of the magnetic field:
\begin{align}\label{MHD1_A18}
	\partial_t\rho =& -\nab\cdot(\rho\vv), \\ \label{MHD2_A18}
	\partial_t\vv =& -\vv\cdot\nab\vv + \va\cdot\nab\va - \frac{1}{\rho}\nab(P+P_M) \nonumber \\
	&- \va\cdot(\dva) + {\bf d_{\nu}} + {\bf f}, \\ \label{MHD3_A18}
	\partial_t\va =& - (\vv-\lambda\jjc)\cdot\nab\va + \va\cdot\nab(\vv-\lambda\jjc) \nonumber\\
	&- \frac{\va}{2}(\dv-\lambda\djc) + \frac{{\bf d_{\eta}}}{\sqrt{\rho}}, \\ \label{MHD4_A18}
	\va\cdot\nab\rho =& -2\rho(\dva),  \\ \label{MHD5_A18}
	\jjc\cdot\nab\rho =& -\rho(\djc),
\end{align}
where $P_M\equiv\rho v_\text{A}^2/2$ is the magnetic pressure and $\rho$ is a time-space variable. The system is once again closed with an isothermal closure. These equations are used to compute the time derivative of the two point correlator
$ R_E  \equiv ~ \frac{\rho}{2}(\vv \cdot \vv'+\va \cdot \va')+\rho e'$,
which ultimately leads to the exact law:
\begin{widetext}
\begin{align} \nonumber
    -2\eA =&~  \frac{1}{2}\boldsymbol\nabla_{\ell}\cdot \lang [(\delta(\rho\vv)\cdot\delta\vv+\delta(\rho\va)\cdot\delta\va + 2\delta e\delta\rho\big]\delta\vv - [\delta(\rho\vv)\cdot\delta\va+\delta\vv\cdot\delta(\rho\va)]\delta\va \rang\\ \nonumber
    +& \lang[R_E'-\frac{1}{2}(R_B'+R_B)+\frac{P_M'-P'}{2}-E^{tot'}](\dv)\rang + \lang[R_E-\frac{1}{2}(R_B+R_B')+\frac{P_M-P}{2}-E^{tot}](\dvp)\rang \\ \nonumber
    +& \lang[(R_H-R_H')-\bar{\rho}(\vv'\cdot\va)+H'+\lambda\delta\rho\frac{\jjc\cdot\va'}{2}](\dva)\rang \\ \nonumber
    +& \lang[(R_H'-R_H)-\bar{\rho}(\vv\cdot\va')+H-\lambda\delta\rho\frac{\jjc'\cdot\va}{2}](\dvap)\rang \\ \nonumber
    +& \frac{1}{2}\lang\big(e'+\frac{v_\text{A}}{2}^{'2}\big)\big[\boldsymbol\nabla\cdot(\rho\vv)\big]+\big(e+\frac{v_\text{A}}{2}^2\big)\big[\boldsymbol\nabla'\cdot(\rho'\vv')\big]\rang - \frac{1}{2}\lang\beta^{-1'}\boldsymbol\nabla'\cdot(e'\rho\vv) + \beta^{-1}\boldsymbol\nabla\cdot(e\rho'\vv') \rang \\ \label{A18}
    +& \frac{1}{2}\boldsymbol\nabla_{\ell}\cdot \lang 2\lambda[(\overline{\rho\jjc\times\va})\times\delta\va-\delta(\jjc\times\va)\times\overline{\rho\va}]\rang + \lambda\lang\frac{R_B-R_B'}{2}(\djc)+\frac{R_B'-R_B}{2}(\djcp)\rang,
\end{align}
\end{widetext}
with the cross helicity $H \equiv \rho(\vv\cdot\nobreak\va)$, its two-points correlator $R_H \equiv \rho(\vv\cdot\va'+\va\cdot\vv')/2$, the correlator for magnetic energy $R_B \equiv \rho\va\cdot\va'/2$ and $\beta^{-1}\equiv v_\text{A}^2/2c_s^2$. Primed variables are obtained by inverting the positions of the primes in the definitions. Just like equation (\ref{F21}), this law is obtained under the assumptions of statistical homogeneity, forcing limited to large scales, time stationarity and infinite Reynolds number. Again, the assumptions on the forcing, dissipation and time stationarity can be dropped by using the general expression obtained directly from the dynamical equation of $\lang R_E + R_E' \rang$:
\begin{equation} \label{fullA18_comp}
2\eA + \partial_t \lang R_E + R_E' \rang + \dA + \fA= 0.
\end{equation}
Here again the same caution as above applies regarding the introduction of $\eA$, whose expression is given by (\ref{A18}). The dissipative and forcing terms $\dA$ and $\fA$ are not explicitly given in \citet{andres18} but are easy to calculate. For the dissipation term, the component stemming from the velocity field and Navier-Stokes equation is the same as in (\ref{DF21}), and the one originating from the magnetic field and induction equation stems from:
\begin{equation}
\partial_t \lang \frac{1}{2}(\rho+\rho') \va \cdot \va' \rang ,
\end{equation}
and reads:
\begin{align} \nonumber
&  \frac{1}{2} \lang \rho \va \cdot \frac{\dd_\eta'}{\sqrt{\rho'}} + \rho \va' \cdot \frac{\dd_\eta}{\sqrt{\rho}} + \rho' \va \cdot \frac{\dd_\eta'}{\sqrt{\rho'}} + \rho' \va' \cdot \frac{\dd_\eta}{\sqrt{\rho}} \rang \nonumber\\
=& \frac{1}{2\mu_0} \lang (\frac{\sqrt{\rho}}{\sqrt{\rho'}}+\frac{\sqrt{\rho'}}{\sqrt{\rho}}) (\bb \cdot \dd_\eta' + \bb' \cdot \dd_\eta) \rang,
\end{align}
which ultimately leads to
\begin{align}
	\dA =&~ -\frac{1}{2}\lang \left(1+ \frac{\rho'}{\rho} \right) \uu' \cdot {\bf d_\nu} + \left(1+\frac{\rho}{\rho'}\right) \uu \cdot {\bf d_\nu'} \rang \nonumber \\ 	-
	&~ \frac{1}{2\mu_0} \lang (\frac{\sqrt{\rho}}{\sqrt{\rho'}}+\frac{\sqrt{\rho'}}{\sqrt{\rho}}) (\bb \cdot \dd_\eta' + \bb' \cdot \dd_\eta) \rang.
\end{align}
The forcing term is identical to the one of law F21, and reads
\begin{align}
	\fA =&~ -\frac{1}{2} \left\langle \left(1+ \frac{\rho'}{\rho} \right) \uu' \cdot {\bf f} + \left(1+\frac{\rho}{\rho'}\right) \uu \cdot {\bf f'} \right\rangle.
\end{align}

        
\section{Numerical methods}

\subsection{Simulation data}

The equations of CHMHD (\ref{MHD1_F21})-(\ref{MHD4_F21}) are solved numerically using the pseudo-spectral code GHOST \citep{gomez05,mininni11} along with a module for solving compressible HMHD flows with the eventual presence of a background magnetic field. Three simulations were run in a cubic periodic box of spatial resolution of $N=1024$ grid points and size $L_0=2\pi$ in all three directions, and all simulations use dimensionless viscosity and magnetic diffusivity of $\nu=\eta=3\times 10^{-4}$. Due to the normalization of the variables, the corresponding Reynolds numbers are simply the inverse of the viscosity and the diffusivity: $R_e=R_m\simeq 3\times 10^{3})$. The ion inertial length $d_i$ is set to $d_i = 0.02 L_0$ for the three Runs. These simulations do not feature any forcing: instead, the kinetic and magnetic fields are set to an initial state built from a superposition of harmonic modes with random phases whose energy in Fourier space is put in a sphere between wave vectors $k_{down}$ and $k_{up}$, following the idea of \citet{pouquet78}. Then, the flow is left to evolve and decay. Varying the intensity of background magnetic field $B_0$ and the initial Mach number $M_S$ allows us to evaluate the influence of these parameters on the dynamics of the system. All aforementioned parameters are reported in Table \ref{sims}.

\begin{table*}[ht]
\centering
\begin{tabular}{ m{4em} m{6em} m{4em} m{4em} m{4em} m{5.5em} m{2.5em} m{2em}} 
 \hline
 Run & Resolution & $B_0$ & $M_S$ & $d_i/L_0$ & $\nu=\eta$ & $k_{down}$ & $k_{up}$ \\ 
 \hline
 I	 & $1024^3$	& $2$	& 0.25	& 0.02 & $3.0\times 10^{-4}$	& 1	& 3 \\ 
 II	 & $1024^3$	& $0$	& 0.25	& 0.02 & $3.0\times 10^{-4}$	& 1	& 3 \\ 
 III & $1024^3$	& $0$	& 0.5	& 0.02 & $3.0\times 10^{-4}$	& 1	& 3 \\ 
 \hline
\end{tabular}
 \caption{List of Runs and their relevant parameters.}
\label{sims}
\end{table*}

\begin{figure}
	\centering
	\subfigure{
	\includegraphics[width=\hsize,trim={0 75 70 75},clip]{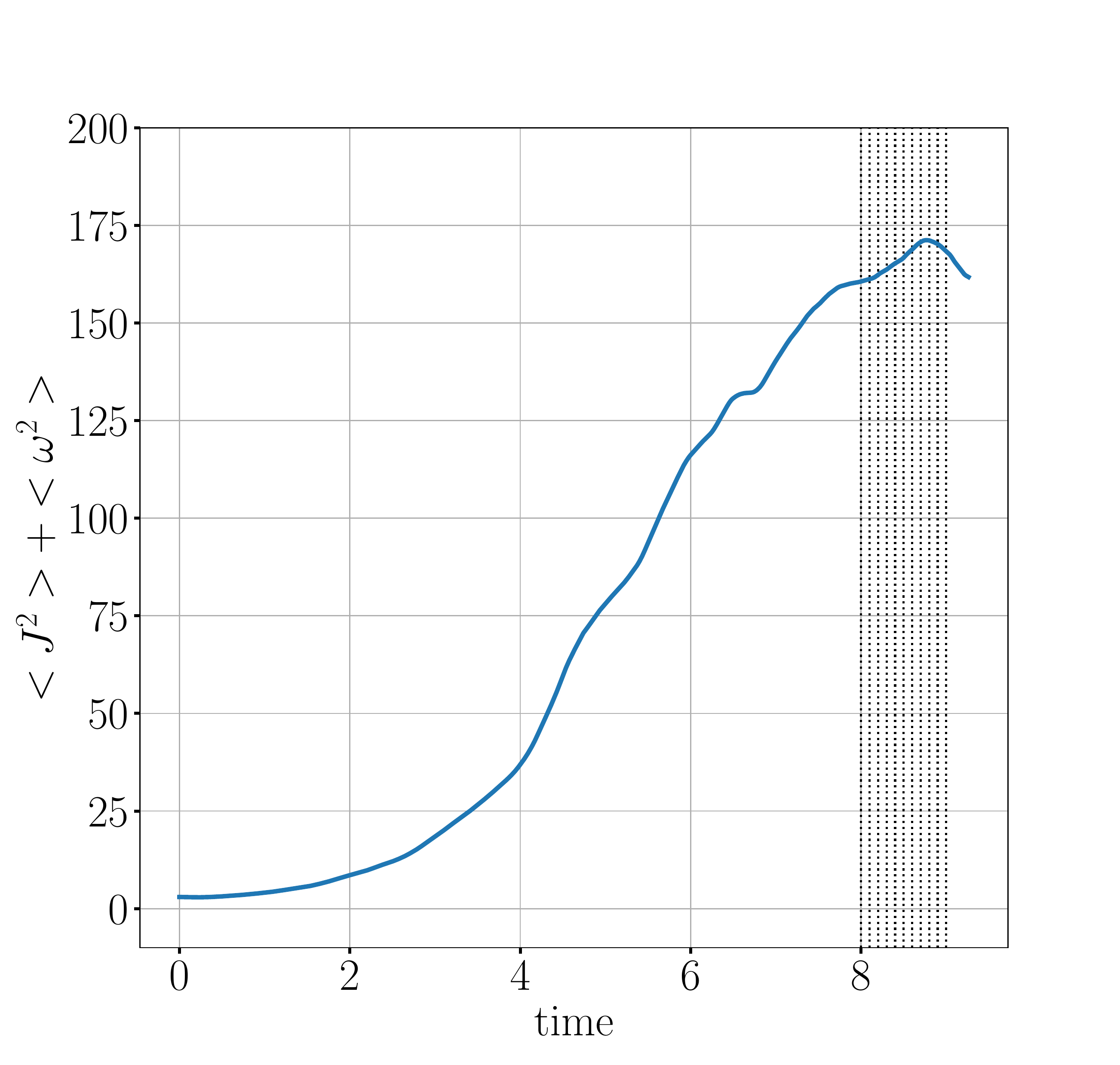}
}
	\subfigure{
	\includegraphics[width=\hsize,trim={0 10 70 85},clip]{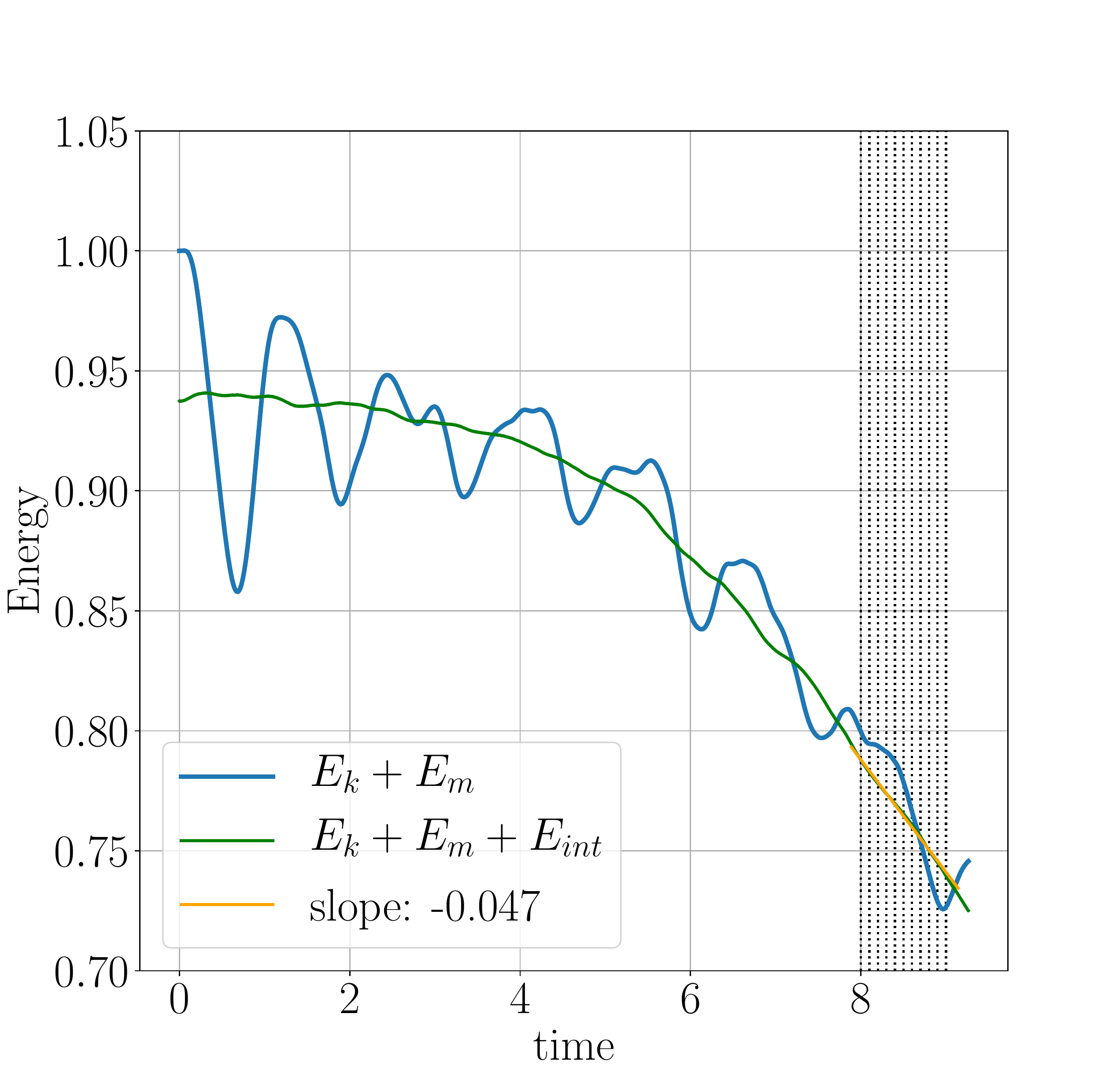}
}
	\caption{Top: Incompressible dissipation as a function of time for Run I. The vertical dotted lines represent the 11 times selected to compute the exact laws. Bottom: Fluctuations of kinetic, magnetic and internal energies (resp. $E_k$, $E_m$ and $E_{int}$) as a function of time for Run I. The total energy (green curve) shows a continuous exchange between kinetic plus magnetic and internal energies. The narrow orange line is a linear fit of the times studied in this paper, and its slope represents the rate of energy loss at these times. }
	\label{energy_I}
\end{figure}

\begin{figure}
	\centering
	\subfigure{
	\includegraphics[width=\hsize,trim={0 75 70 75},clip]{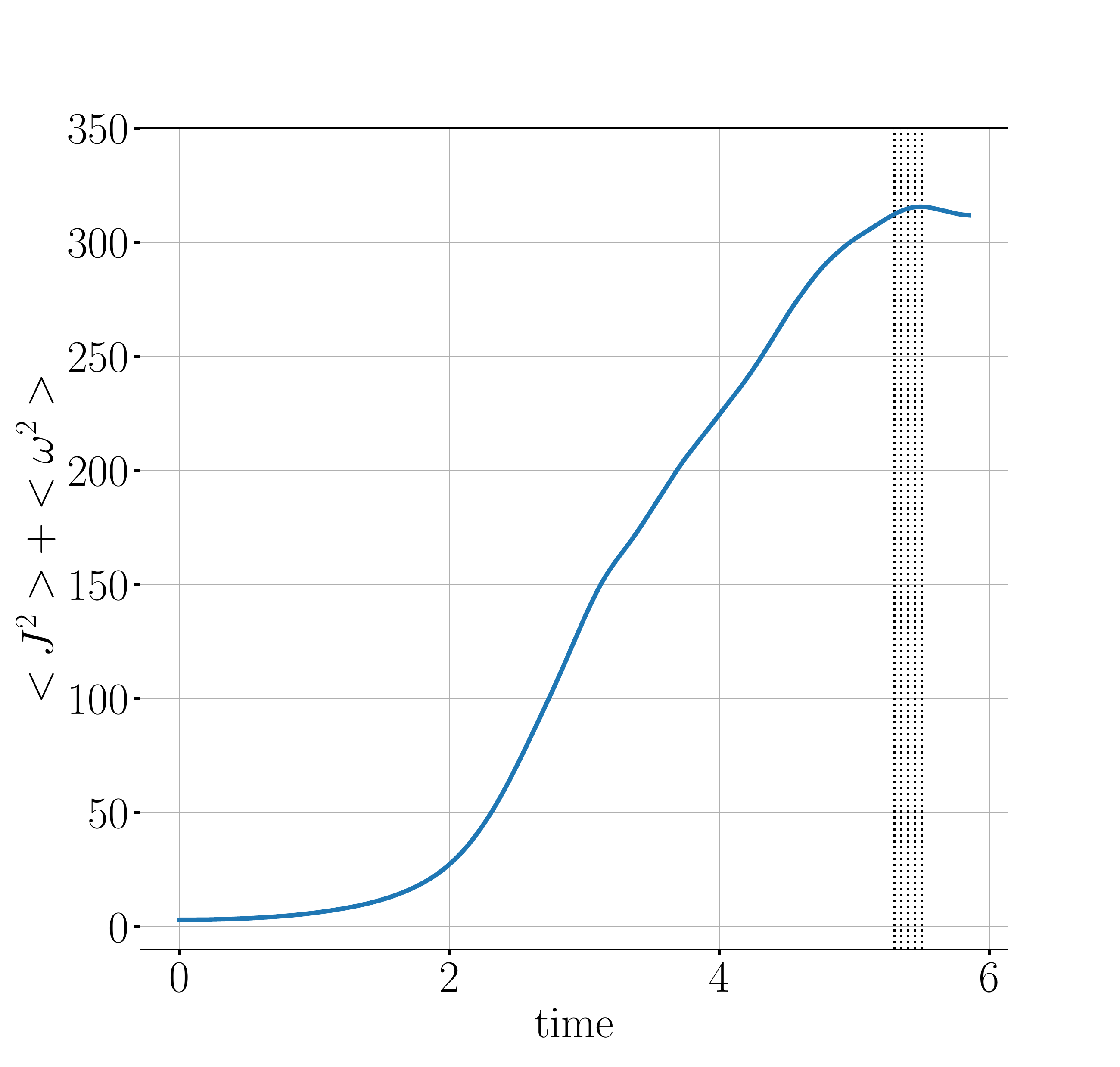}
}
	\subfigure{
	\includegraphics[width=\hsize,trim={0 10 70 75},clip]{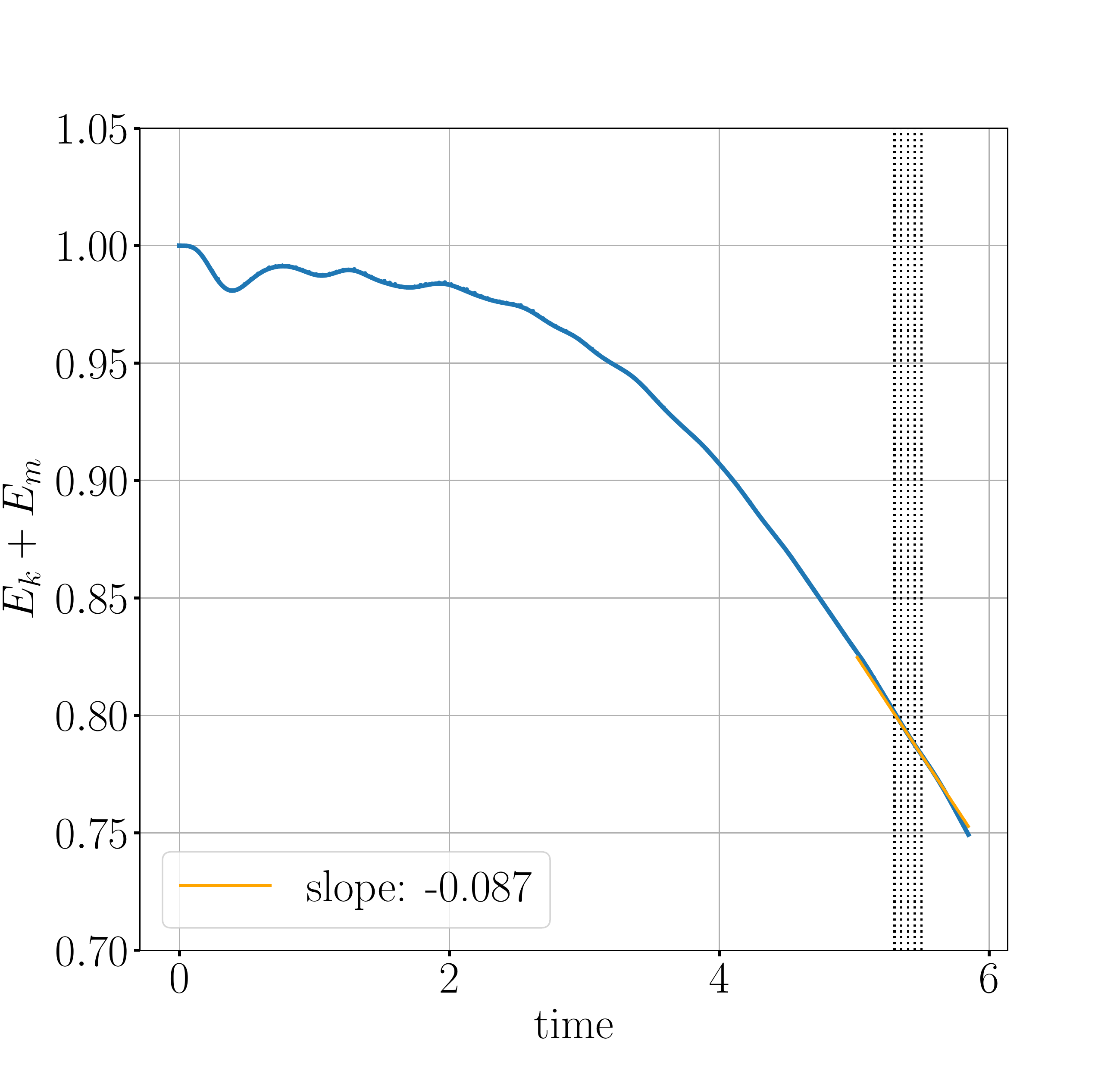}
}
	\caption{Top: Incompressible dissipation as a function of time for Run II. The vertical dotted lines represent the 5 times selected to compute the exact laws. Bottom: Fluctuations of kinetic plus magnetic energy as a function of time for Run II. The narrow orange line is a linear fit of the times studied in this paper, and its slope represents the rate of energy loss at these times. }
	\label{energy_II}
\end{figure}

As the simulations are free-decay, instead of looking for a stationary regime, we wait for the simulations to reach a maximum of dissipation, indicating that turbulence has had enough time to fully develop, and lead our study on times selected around this moment. Around these times the sum of the kinetic and magnetic energies of the system, $E_k+E_m=\rho u^{2}/2 + B^{2}/(2\mu_{0})$ is expected to be decreasing steadily. Figures \ref{energy_I} and \ref{energy_II} show these dissipation and energy for Runs I and II (Run III exhibits an almost identical behavior to Run II). The steady decline of the energy is indeed observed for Run II, yet Run I shows oscillations on top of the general behavior. These oscillations, as shown in Fig. \ref{energy_I}, match fluctuations of the internal energy and are thus thought to be a consequence of exchanges between the kinetic plus magnetic energy and internal energy, as was already reported in \citet{yang21}, and initiated by the presence of waves. Using a linear fit, one can estimate the energy dissipation rate at the selected times: for Run I it is estimated at $\sim -0.047$ and for Runs II and III at $\sim -0.087$. If the energy cascade rate is indeed representative of the energy transferred to and dissipated at small scales, its amplitude should match these estimates for the various simulations. We will return to this point in the next section.

\subsection{Methods of calculation} \label{methods}

To compute all terms from equations (\ref{fullF21_comp}) and (\ref{fullA18_comp}) with large-enough statistics we use two different numerical schemes depending on whether there is a background magnetic field in the simulation or not. Both originate from the discrete decomposition of space proposed by \citet{taylor03}: increment vectors $\el$ are selected along 73 directions defined by base vectors connecting two points of the grid. Increments are taken as multiples of these base vectors so that both ${\textbf r}$ and ${\textbf r}'={\textbf r}+\el$ lie on known grid points, allowing for a well mapping of space without having to interpolate the 3D data, which would be a very time-consuming process. Such a decomposition has already been successfully used to compute the two-point correlation functions in simulation data of compressible MHD turbulence \citep{andres18b}.

All source terms (i.e. terms that do not appear as divergences along the increment vector) can be computed directly as long as the increment vector is known. We only need to proceed to the ensemble average, which in our case is taken on the full simulation domain: 
\begin{equation}
	\lang \xi \rang = \sum_{\textbf r} \frac{\xi({\textbf r})}{N^3},
\end{equation}
where $N=1024$ is the number of the grid points.  

For flux terms (i.e. terms that appear as divergences along the increment vector, of the form $\nab_{\el} \cdot {\textbf F}$) we make the assumption that our system is isotropic. In this situation we naturally use spherical coordinates, in which $\el$ is defined as $\el = (\ell,\phi,\theta)$ and the derivative operator $\nab_{\el}$ reduces to $\nab_{\el} \cdot \lang {\textbf F} \rang = \frac{1}{\ell^2}\partial_{\ell} [\ell^2 \lang F_{\ell} \rang(\ell)]$. Thus, for a given increment vector $\el$ we only have to compute the projection of the vectorial flux ${\textbf F}$ on the direction of $\el$, which writes: 
\begin{equation}
	\lang F_{\ell} \rang (\ell,\phi,\theta) = \lang cos(\phi)sin(\theta)F_x + sin(\phi)sin(\theta)F_y + cos(\theta)F_z \rang.
\end{equation}
These projections are then averaged at fixed $\ell$:
\begin{equation}
	\lang F_{\ell} \rang (\ell) = \sum_{\phi,\theta} \frac{\lang F_{\ell} \rang (\ell,\phi,\theta)}{n_{dir}}, 
\end{equation}
where $n_{dir}=73$ refers to the number of different directions taken for $\el$.

Note that, the isotropy assumption stands for Runs II and III in which $\bb_0=0$. A similar method based on the assumption of a symmetry of revolution along the axis of $\bb_0$ was also used to study Run I. While this method is \textit{a priori} more suited for the study of simulations with $\bb_0 \neq 0$, the isotropic decomposition of the data ultimately provided better results even on Run I, and is thus the only one used in this paper. A more detailed discussion on this point is given in the Appendix.

\subsection{Applied calculation}

Using the method described above we compute the various terms of equations (\ref{fullF21_comp}) and (\ref{fullA18_comp}). Note however that since the present simulations are free-decay the forcing terms appearing in those equations are identically zero. Therefore, the latter reduce to  
\begin{equation} \label{fullF21_comp2}
2\eF + \partial_t \lang E^{tot} \rang - \partial_t \lang S \rang + \dF + \dFm  = 0
\end{equation}
for the F21 model and to 
\begin{equation} \label{fullA18_comp2}
2\eA + \partial_t \lang R_E + R_E' \rang + \dA = 0.
\end{equation}
for the A18 model. 

To lead the calculations we retained a number of snapshots for each Run: 11 snapshots evenly spaced in time for Run I (that present fluctuations on the incompressible energy and thus require a time average) between 8 and 9 turnover times, and 5 evenly spaced snapshots for Runs II and III between 5.3 and 5.5 turnover times and 5.6 and 5.8 turnover times respectively.
All time derivatives are obtained by using a five-points finite differences method:
\begin{align} \label{5pt_stencil}
	f'(t) \approx \frac{f(t-2h)-8f(t-h)+8f(t+h)-f(t+2h)}{12h},
\end{align} 
where $h$ represents the time step between two selected snapshots. For Run I, time derivatives (e.g. $\partial_t \lang S \rang$) are calculated on all possible subsets of 5 consecutive snapshots among the initial 11, for a total of 7 calculations (one using times 8 to 8.4, one using times 8.1 to 8.5 etc.), then the resulting derivatives are averaged over time. For all other terms (e.g. $\varepsilon_{F21}$) we compute the time average over all 11 snapshots. For Runs II and III, the derivatives are calculated on the 5 snapshots retained using the same derivation method and all other terms are calculated only on the central snapshot, respectively at times 5.4 and 5.7. Averaging these terms over the five snapshots was found to bring no change to the results (not shown).

\section{Calculation of the energy cascade rate in the inertial range}

\begin{figure}
	\centering
	\includegraphics[width=\hsize,trim={40 0 80 40},clip]{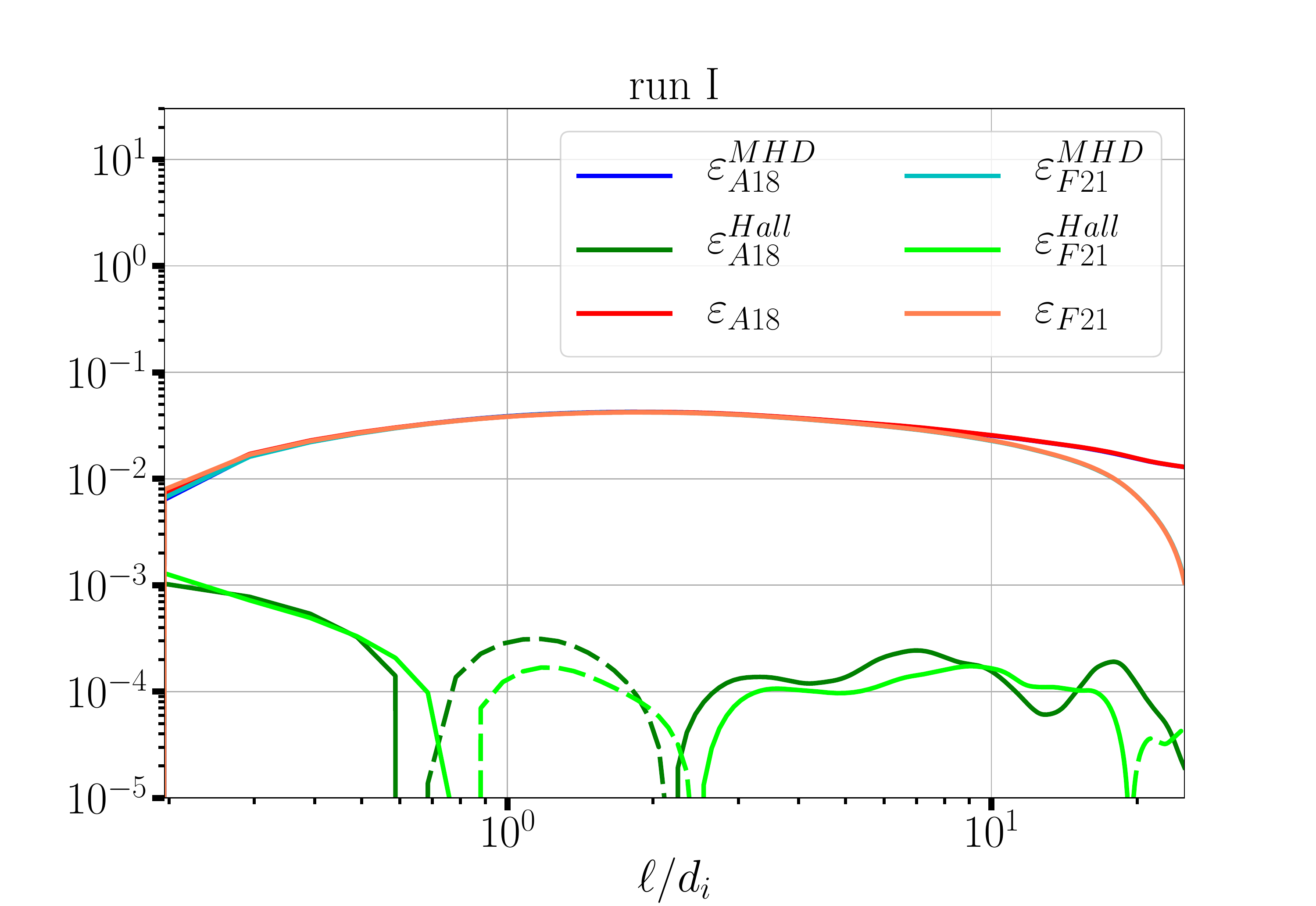}
	\includegraphics[width=\hsize,trim={40 0 80 40},clip]{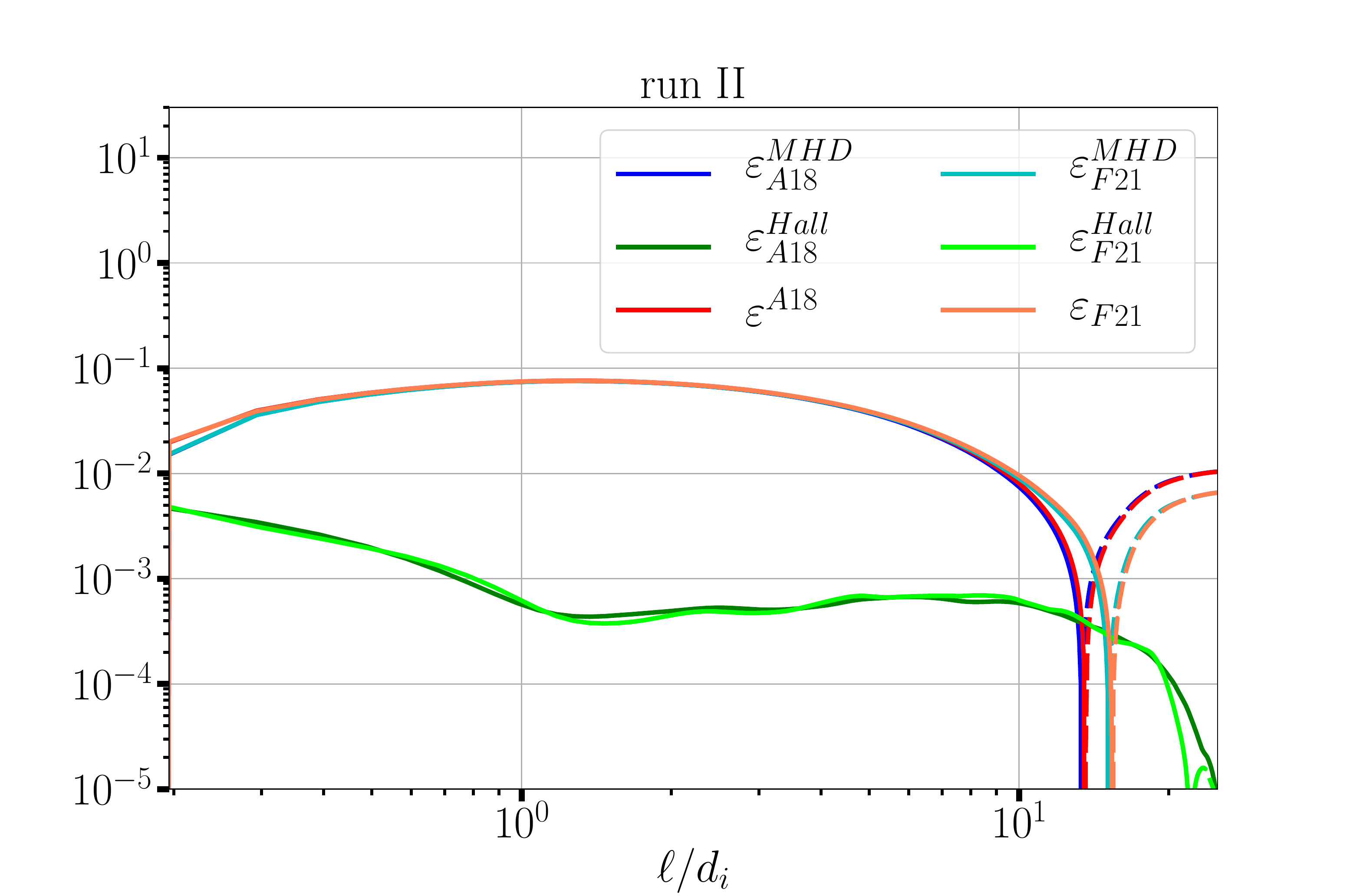}
	\caption{Comparison between the different components of the cascade rate given by the two models of \citet {ferrand21} (Eq. \ref{F21}) and \citet{andres18} (Eq. \ref{A18}) for Run I (top) and Run II (bottom). Plain lines represent positive values whereas dashed lines represent negative ones.}
	\label{runs_comp}
\end{figure}

We first study the energy cascade rates obtained for laws F21 and A18 classically obtained under the full assumptions of space homogeneity, time stationarity and infinite Reynolds number, i.e. equations (\ref{F21}) and (\ref{A18}) respectively. For both laws, energy cascade rates are broken down into a Hall component $\varepsilon^{Hall}$ and a MHD component $\varepsilon^{MHD}$, which are made respectively of the terms in factor of $\lambda$ and of all the remaining terms from equations (\ref{F21}) and (\ref{A18}). A comparison of the cascade rates provided by the two exact laws is shown in Fig. \ref{runs_comp} for Runs I and II. One can observe that the two models yield closely similar components (MHD and Hall) of the cascade rate nearly at all scales. This validates numerically the equivalence of the two exact laws in the inertial range as anticipated in \citet{ferrand21}. Note that in both Run I and Run II the value of the cascade rate in the inertial range, centered around $\ell/d_i \simeq 2$, roughly matches the energy dissipation rate estimated through the linear fit on the energy (respectively $\sim -0.047$ and $\sim -0.087$, see Figs. \ref{energy_I} and \ref{energy_II}), suggesting that the energy cascade approximated in the inertial range is representative of the dissipation in the system. The difference in the cascade rate values between the two runs is due to differences in the initial/driving amplitude of fluctuations, which is the lowest for Run I with $B_0=2$. 

\begin{figure}
	\centering
	\includegraphics[width=\hsize,trim={40 0 80 70},clip]{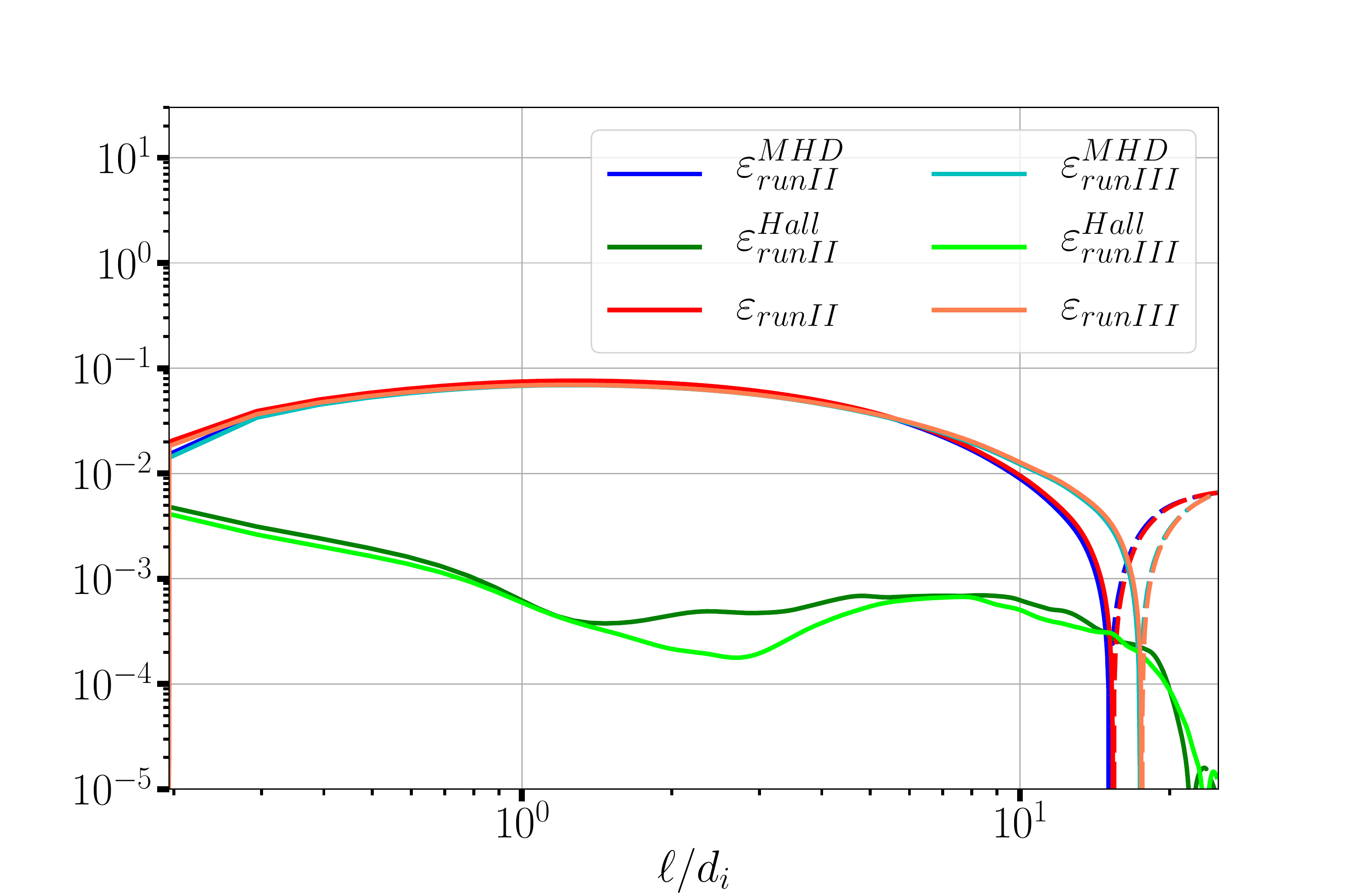}
	\caption{Comparison between the cascade rate given by model \citet {ferrand21} (Eq. \ref{F21}) for two different Mach numbers: $M_S=0.25$ (Run II) and $M_S=0.5$ (Run III)}
	\label{mach}
\end{figure}

Another question that can be addressed regarding the cascade rate in the inertial range is its sensitivity (or not) to the turbulent sonic Mach number. To do so we compare in Fig. \ref{mach} the results of Runs II and III that correspond to the initial Mach numbers $M_S=0.25$ and $M_S=0.5$, respectively. We observe that, overall, the cascade rate are very close to each other at all scales, and appear to be similar among all three Runs, indicating that the increase in the Mach number from $M_S=0.25$ to $M_S=0.5$ does not bring significant changes to the total dynamics of the system. This result is in agreement with the findings in \citet{andres18b} for compressible MHD turbulence who already reported that, for Mach numbers up to $M_S=0.5$, purely compressible components of the exact law remain negligible in comparison to the flux terms, which only slightly deviate from their incompressible counterparts. Note that such conclusions only hold \textit{a priori} for subsonic regimes: supersonic turbulent flows can develop a dominant compressible (source-like) component of the energy cascade over the traditional flux driven one \citep{ferrand20}. 

At this point, an important remark can be made: all the energy cascade rates reported in Figs. \ref{runs_comp} and \ref{mach} have a relatively low amplitude Hall component, which never becomes dominant with regard to the MHD component in contrast with results reported previously from 3D CGL simulations \citep{ferrand21b} (i.e., simulations using a closure with an anisotropic pressure tensor introduced by Chew, Goldenberg and Low \citep{CGL56}, hence the acronym). An explanation can be given: the Hall effect remains too weak in our simulations such that the dissipation inhibits the energy cascade before its Hall component becomes dominant.

\begin{figure}
	\centering
	\includegraphics[width=\hsize,trim={0 0 80 70},clip]{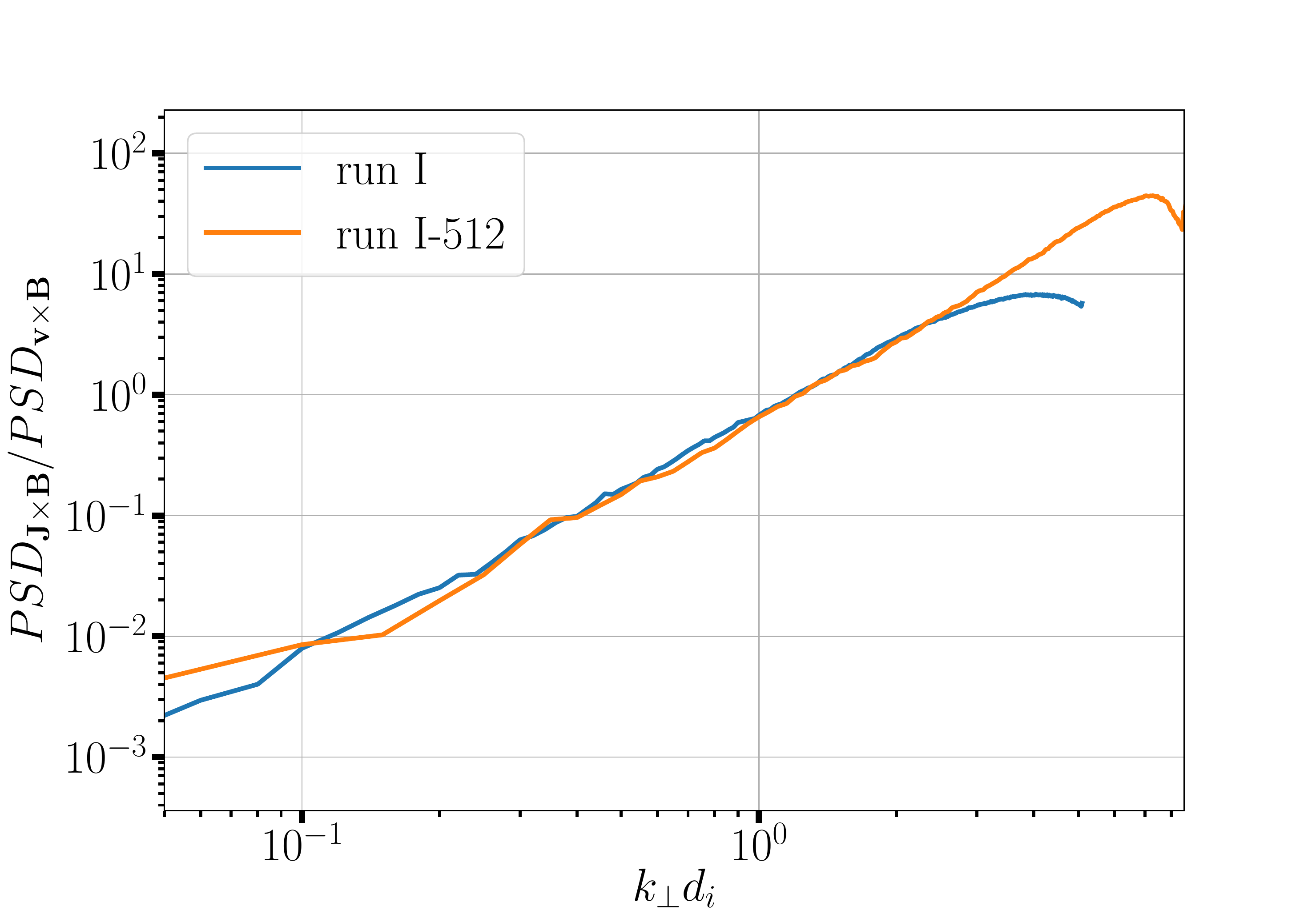}
	\caption{Ratio of the power spectrum density of the Hall to the ideal ($\vv \times \bb$) components of the electric field for Run I and the lower resolution Run I-512. Note that the window has been set to only show the scales spanned by the spectrum of run I-512 ; the one for run I extends further towards smaller values of k.}
	\label{E_spectra}
\end{figure}

To test this hypothesis we ran an additional simulation (Run I-512) that is akin to Run I but with a lower resolution $N=512$, a slightly higher dissipation $\nu=\eta=8.0\times 10^{-4}$ and an ion inertial length $d_i = 0.05/L_0$. $L_0$ and $M_S$ are kept unchanged. Reducing the resolution while increasing the value of $d_i$ (despite the slight increase of the dissipation) allows for increasing the size of the sub-ion range, which should in turn increase the importance of the Hall effect at the smallest scales available. A first measure of this can be obtained by looking at the power spectrum of the electric field, defined by the reduced Ohm's law as (omitting the resistive term):
\begin{equation} \label{Ohm}
	{\bf E} = -\vv \times \bb + \frac{1}{nq_e}\jj \times \bb,
\end{equation}
where $n$ is the particle density. Fig. \ref{E_spectra} shows the ratio of the power spectrum density (PSD) of the Hall to the ideal components of the electric field for Run I and the lower resolution Run I-512. As expected, we observe both an increase in the amplitude of the Hall electric field and its extension to smaller scales for Run I-512 in comparison with Run I.

\begin{figure}
	\centering
	\includegraphics[width=\hsize,trim={45 0 85 40},clip]{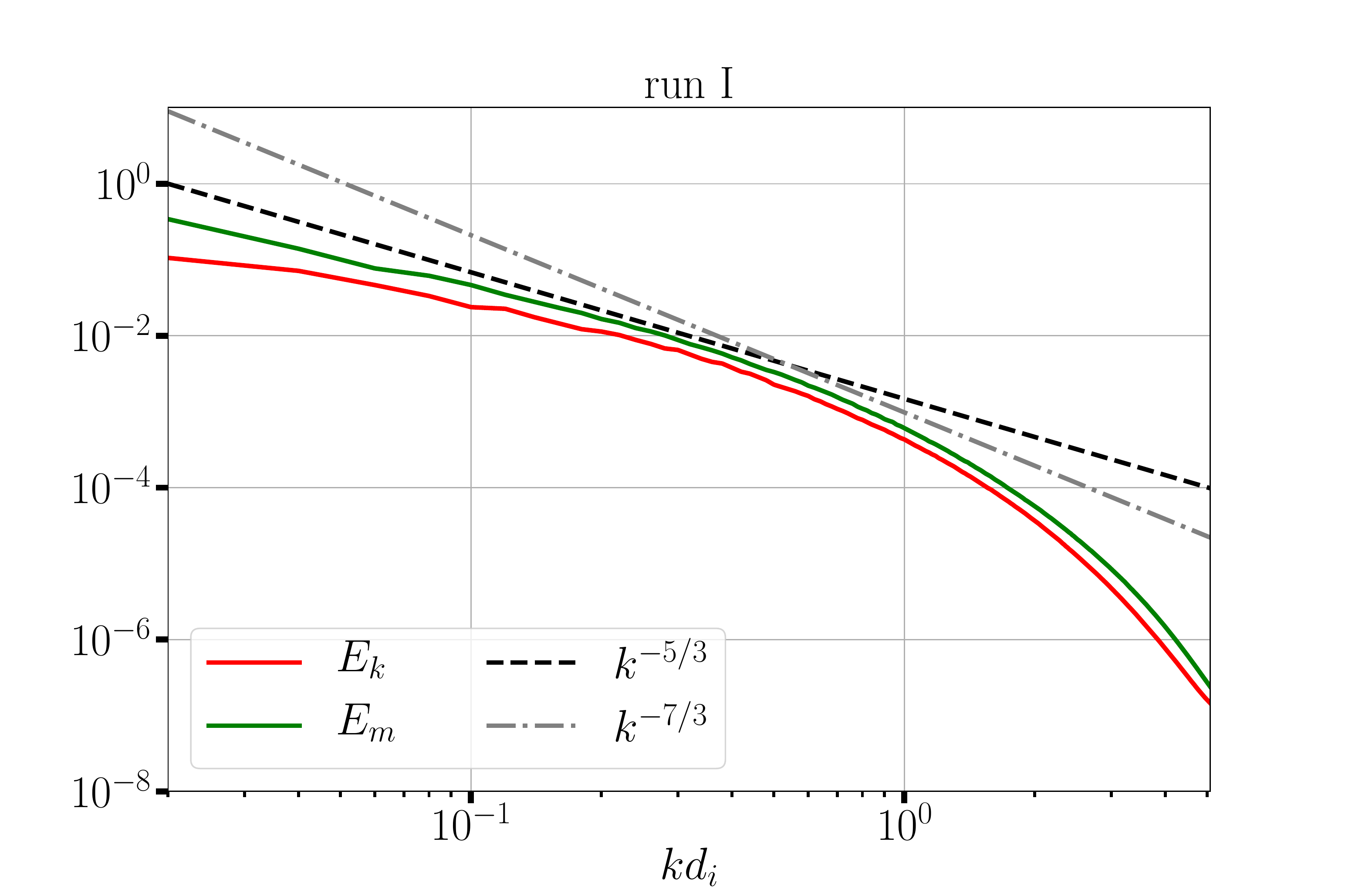}
	\includegraphics[width=\hsize,trim={45 0 85 20},clip]{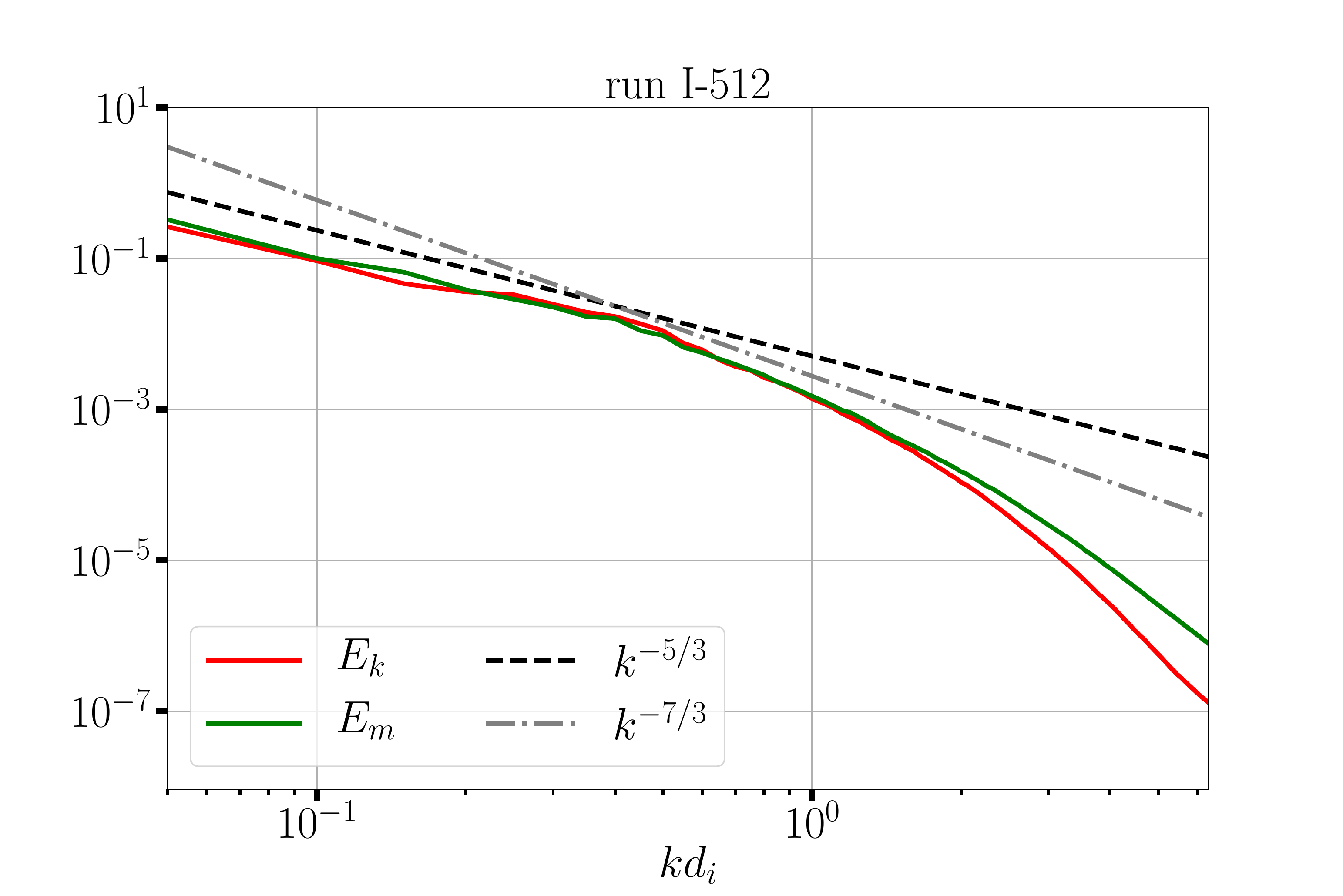}
	\caption{Kinetic and magnetic power spectra (in red and green respectively) for run I (top) and run I-512 (bottom). Reference power laws of slopes -5/3 and -7/3 are represented by a black dashed line and a grey dash-dot line respectively.}
	\label{spectra}
\end{figure}

\begin{figure}
	\centering
	\includegraphics[width=\hsize,trim={40 0 80 70},clip]{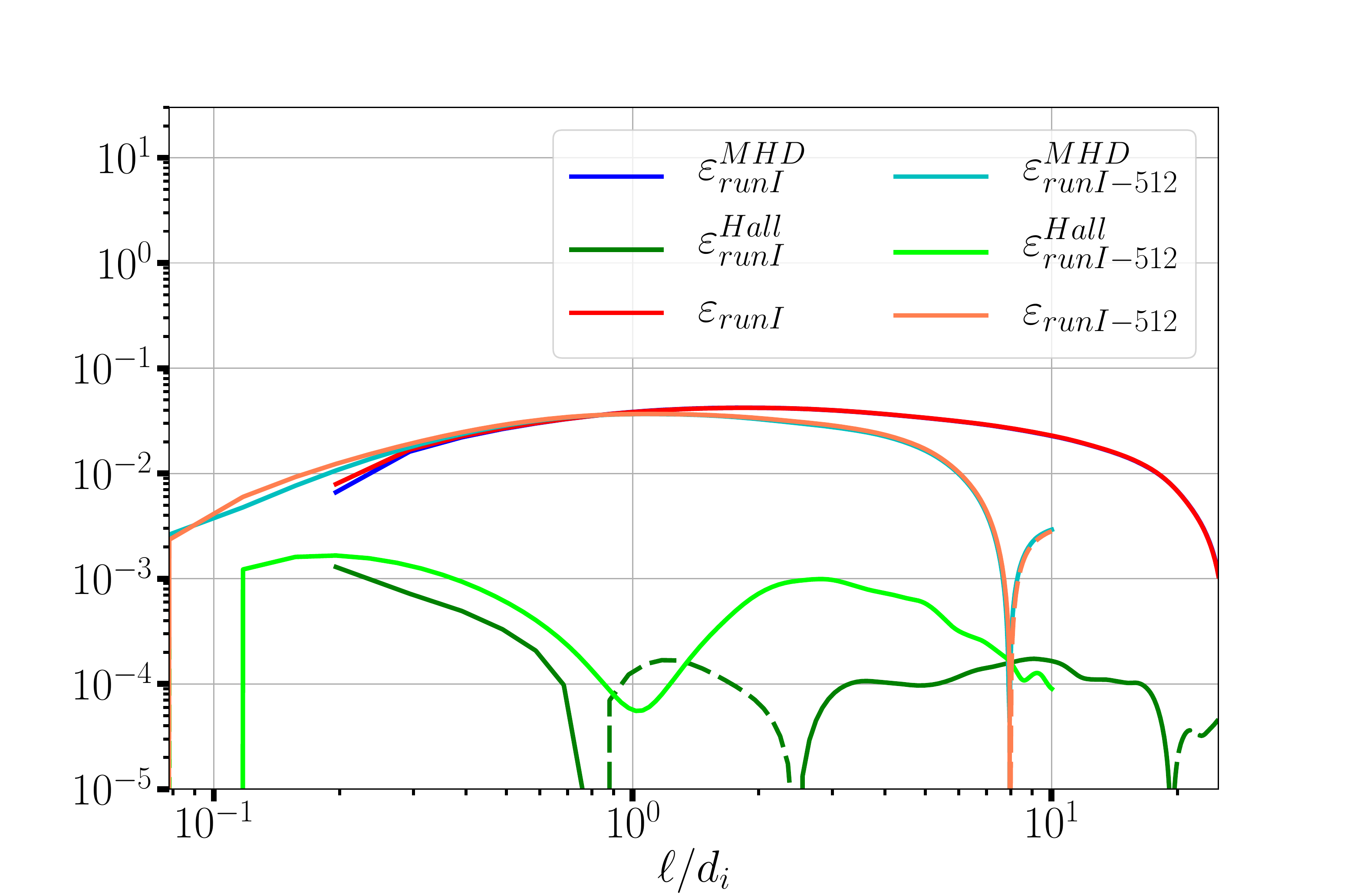}
	\caption{Comparison between the cascade rate given by model \citet {ferrand21} (Eq. \ref{F21}) for Run I and Run I-512.}
	\label{rerunI}
\end{figure}

Then, we look at how the increased Hall effect witnessed in Run I-512 is reflected on the other spectra and the energy cascade rate. Fig. \ref{spectra} represents the kinetic and magnetic power spectra of the simulations for both Run I and Run I-512. The slope of the magnetic spectrum provides a means to pinpoint the MHD inertial range (slope of -5/3) and the Hall range (slope of -7/3). It appears that, in both runs, the MHD range corresponds to a wide range of scales above the ion scale, as indicated by the relatively steady -5/3 scaling. Typically, for Run I, the -5/3 scaling roughly spans $kd_i \in [0.1,0.7]$, which corresponds to $\ell/d_i \in [1.4,10]$ and is in line with the results reported in Fig. \ref{runs_comp} for the energy cascade rate. For the Hall range however, it appears that the spectra quickly drop below the ion scale, straying away from the -7/3 scaling. The drop is less noticeable in Run I-512,that features a stronger Hall effect, still no clear -7/3 scaling can be witnessed at sub-ion scales. However, a clearer knee with a change in slope can be observed in the spectral scaling of this run at wave numbers slightly smaller than the inverse of the ion scale. These observations suggest that, as expected, the Hall regime enters in competition with the increasing dissipation at small scales. Similarly, when looking at the energy cascade rates that are reported in Fig. \ref{rerunI}, we observe little enhancement in the amplitude of the Hall component of Run I-512, but it still does not dominate the cascade at sub-ion scales despite the Hall effect being stronger.

These results call for a further increase in the potency of the Hall effect and/or the  introduction of hyperviscosity and hyperdiffusivity in the GHOST code, which would push the dissipation to the smallest possible scales (under the assumption that the mechanisms responsible for energy dissipation in a plasma act at scales much smaller than the ion inertial length, or are more localized in spectral space than the usual viscosity and Ohmic dissipation). The latter solution was indeed used in the CHMHD-CGL simulations where the Hall cascade was found to dominate below the ion inertial length \citep{ferrand21b}. The former however was not reasonably feasible in this study due to computational constraints. Indeed, for the simulation code to run properly, and as a result of the explicit time resolution of the dispersion of whistlers in Hall MHD, one needs to set an adequate value for the time step of the calculations. This time step has, among other constraints resulting from the Courant-Friedrichs-Lewy condition, an upper bound that roughly behaves as $\sim (dx)^2/(d_i B)$ where $dx$ is the grid resolution. Thus, increasing $d_i$ too much while keeping the grid separation small drastically tightens the constraint on the time step, up to a point where the computational cost becomes prohibitive. Ultimately, at this point, hyperdissipation appears as the best available solution.

\section{Exact laws beyond the inertial range and time stationarity}

As we already stated, switching from a forced turbulence model to a free-decay model invalidates the stationarity hypothesis used to derive the final expression of the compressible exact laws, namely equations (\ref{F21})-(\ref{A18}). Thus, we make use in this section of equations (\ref{fullF21_comp2}) and (\ref{fullA18_comp2}) describing the general exact laws F21 and A18, free from these hypotheses and of the presence of an external forcing. All the terms in those equations are computed for the three Runs and are displayed in figures \ref{runs_A18} and \ref{runs_F21}, along with their sums that, according to the aforementioned equations, should amount to zero.

\subsection{Full equation for law A18}

\begin{figure}
	\centering
	\includegraphics[width=\hsize,trim={40 35 70 40},clip]{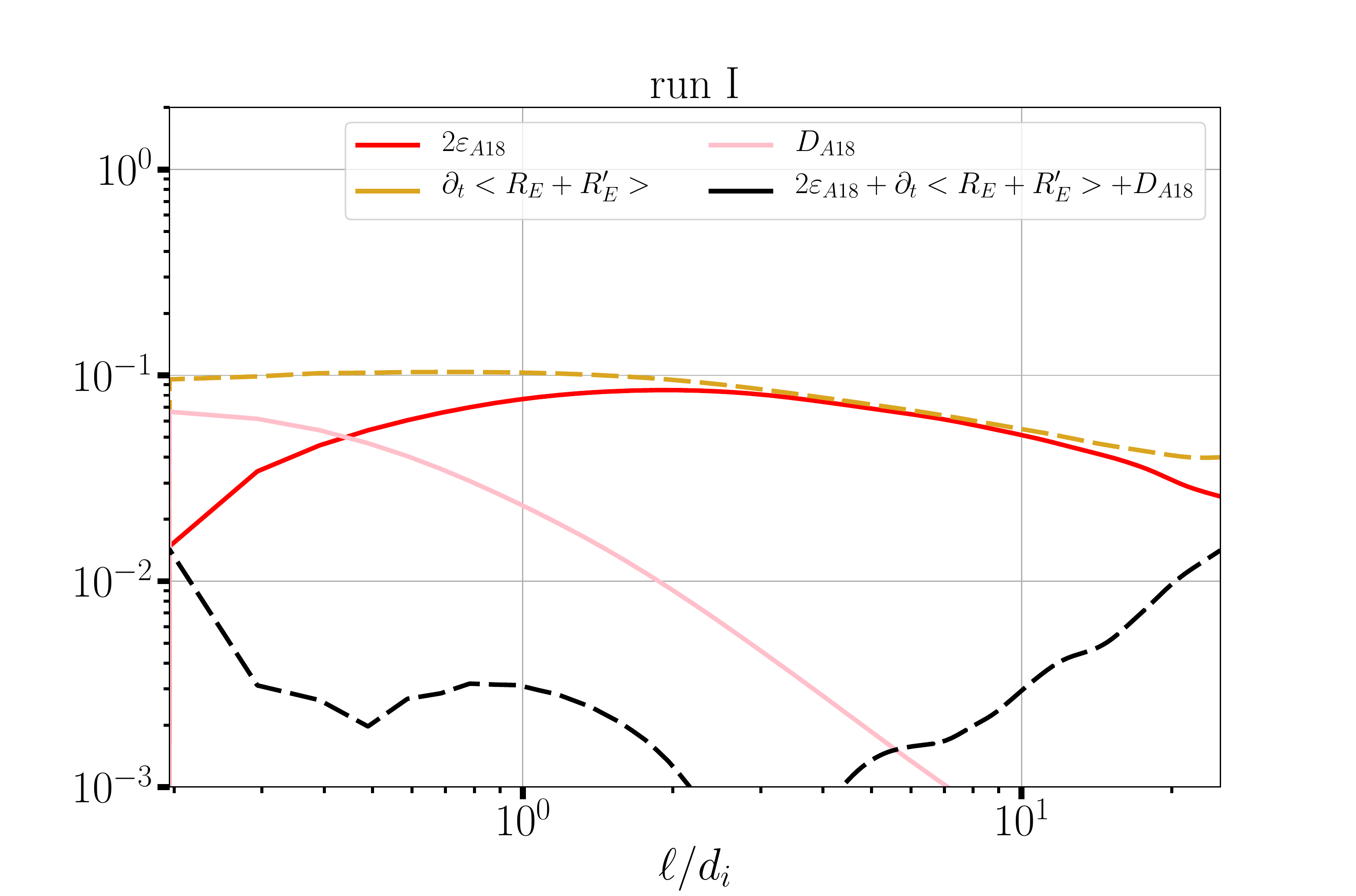}
	\includegraphics[width=\hsize,trim={40 35 70 40},clip]{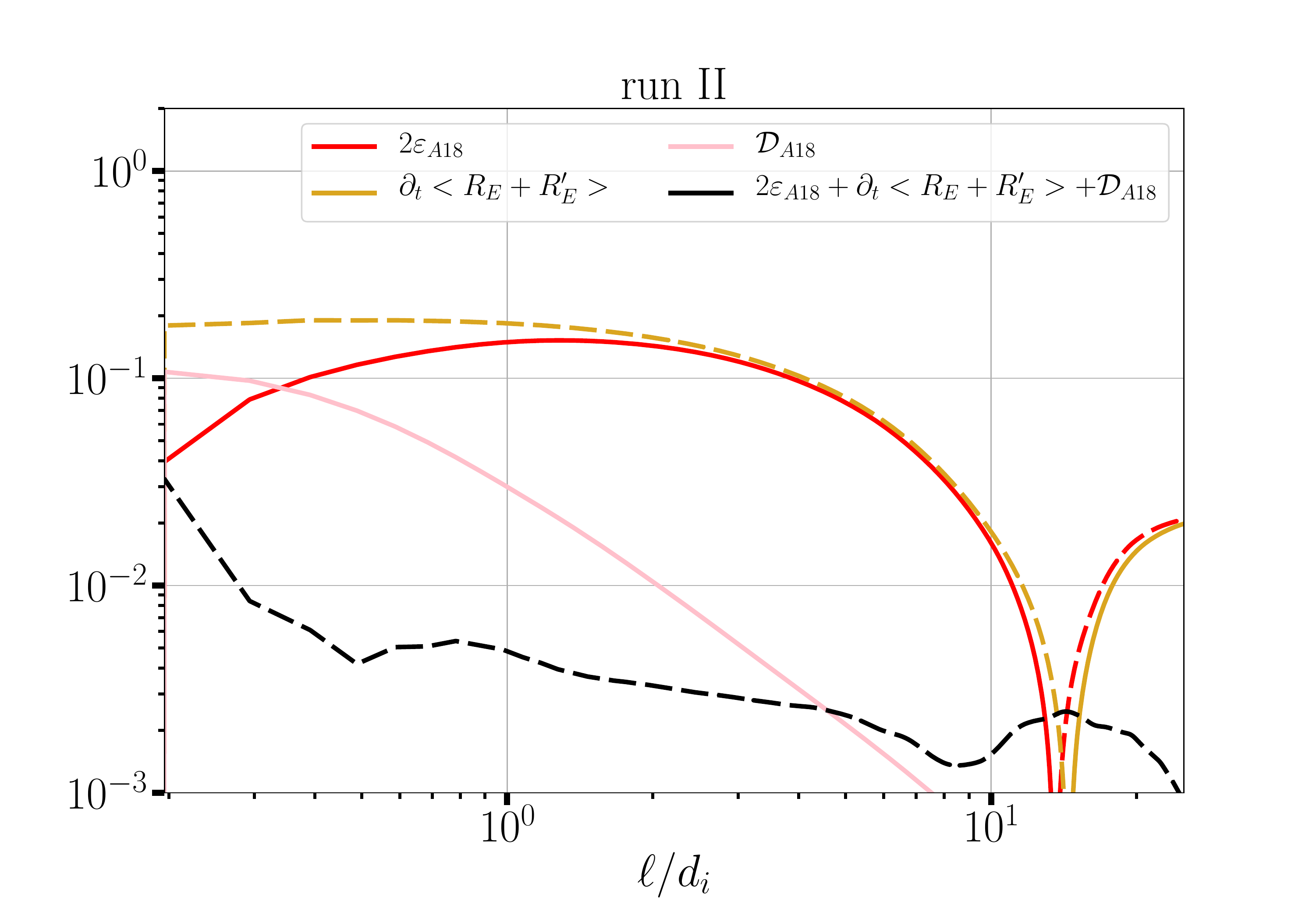}
	\includegraphics[width=\hsize,trim={40 0 70 40},clip]{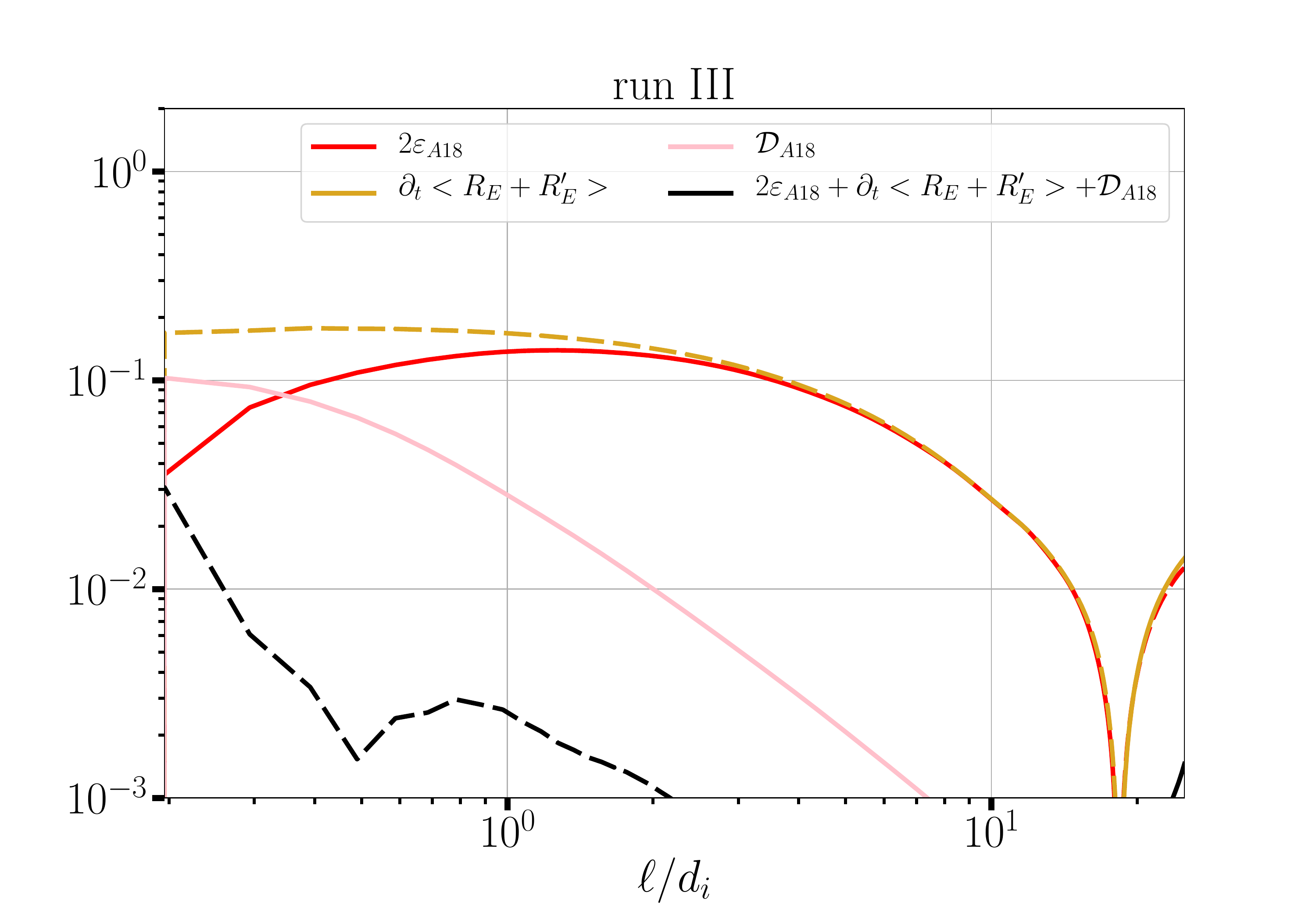}
	\caption{Calculation of equation (\ref{fullA18_comp2}) (black line) and its components for all three Runs. Plain lines represent positive values whereas dashed lines represent negative ones.}
	\label{runs_A18}
\end{figure}

Equation (\ref{fullA18_comp2}) gives a very similar behavior for all three Runs. The sum of all terms, which is supposed to be zero, lies around 1 to 1.5 orders of magnitude below all other terms, which is reasonable given the statistical and discrete nature of the numerical calculations led here. Theoretical limit cases can be easily evaluated here: at large and intermediate scales the dissipation term should be negligible as it represents a mean of uncorrelated terms, resulting in the equation:
\begin{equation} \label{fullA18_ls}
2\eA + \partial_t \lang R_E + R_E' \rang = 0.
\end{equation}
This relation is overall well verified by Run II and III at large scales as reflected by the matching between the red and yellow curves in Fig. \ref{runs_A18}. However, this does not seem to be the case in Run I and the black curve rises at large scales as a result. The reason for this odd behavior is not fully understood yet. At small scales the dissipation is expected to kick in and take energy away from the cascade $\eA$. Also, at small scales we have $\textbf{x} \to \textbf{x}'$, and therefore $\partial_t \lang R_E + R_E' \rang \to 2\partial_t\lang E^{tot} \rang$, resulting in the equation:
\begin{equation} \label{fullA18_ss}
2\eA + 2\partial_t\lang E^{tot} \rang + \dA = 0.
\end{equation}
This equation too is verified in all three Runs, since the black curve in Fig. \ref{runs_A18} that represents a measure of any departure from the perfect fulfillment of the equation is at least 1 order of magnitude lower than the other components of the equation.

The physical interpretation of Eq. (\ref{fullA18_comp2}) is rather simple and can be summed up as follows: in free-decay simulations the term $ \partial_t \lang R_E + R_E' \rang$ in the A18 model plays the role of a ``forcing'', i.e., $-\partial_t \lang R_E + R_E' \rang \equiv \fF$ that inputs energy into the system at each time step. This reservoir of energy is then split into a cascade component with a rate $\eA$ and a dissipation one with a rate $\dA$. At large scale, since $\dA \to 0$ all the energy is almost entirely injected in the cascade towards smaller scales. The sum of all terms, which should be equal to zero, can be seen as an estimation of the error induced in the calculations of time derivatives by the fluctuations of energy.

\subsection{Full equation for law F21}

\begin{figure}
	\centering
	\includegraphics[width=\hsize,trim={40 35 70 40},clip]{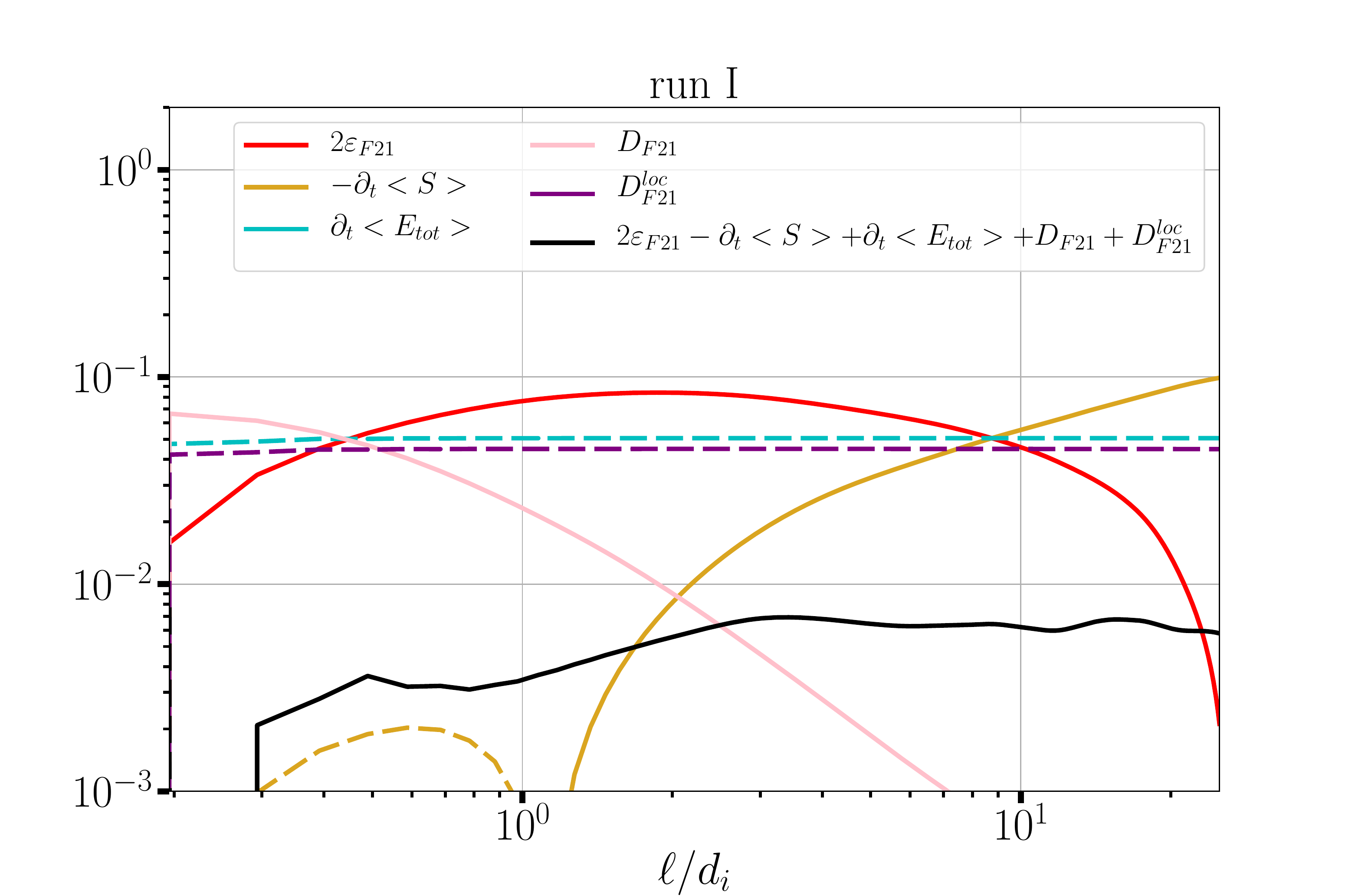}
	\includegraphics[width=\hsize,trim={40 35 70 40},clip]{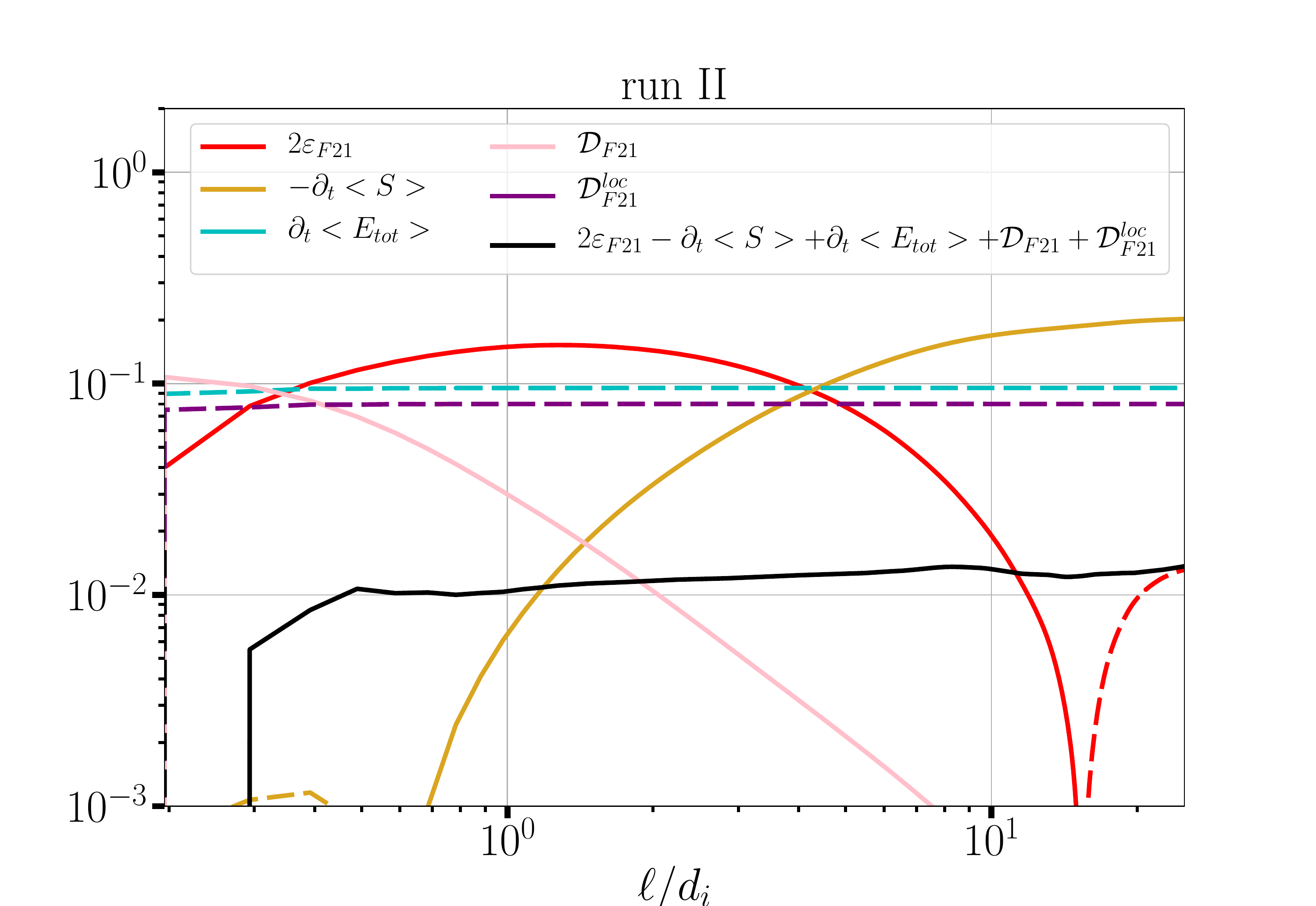}
	\includegraphics[width=\hsize,trim={40 0 70 40},clip]{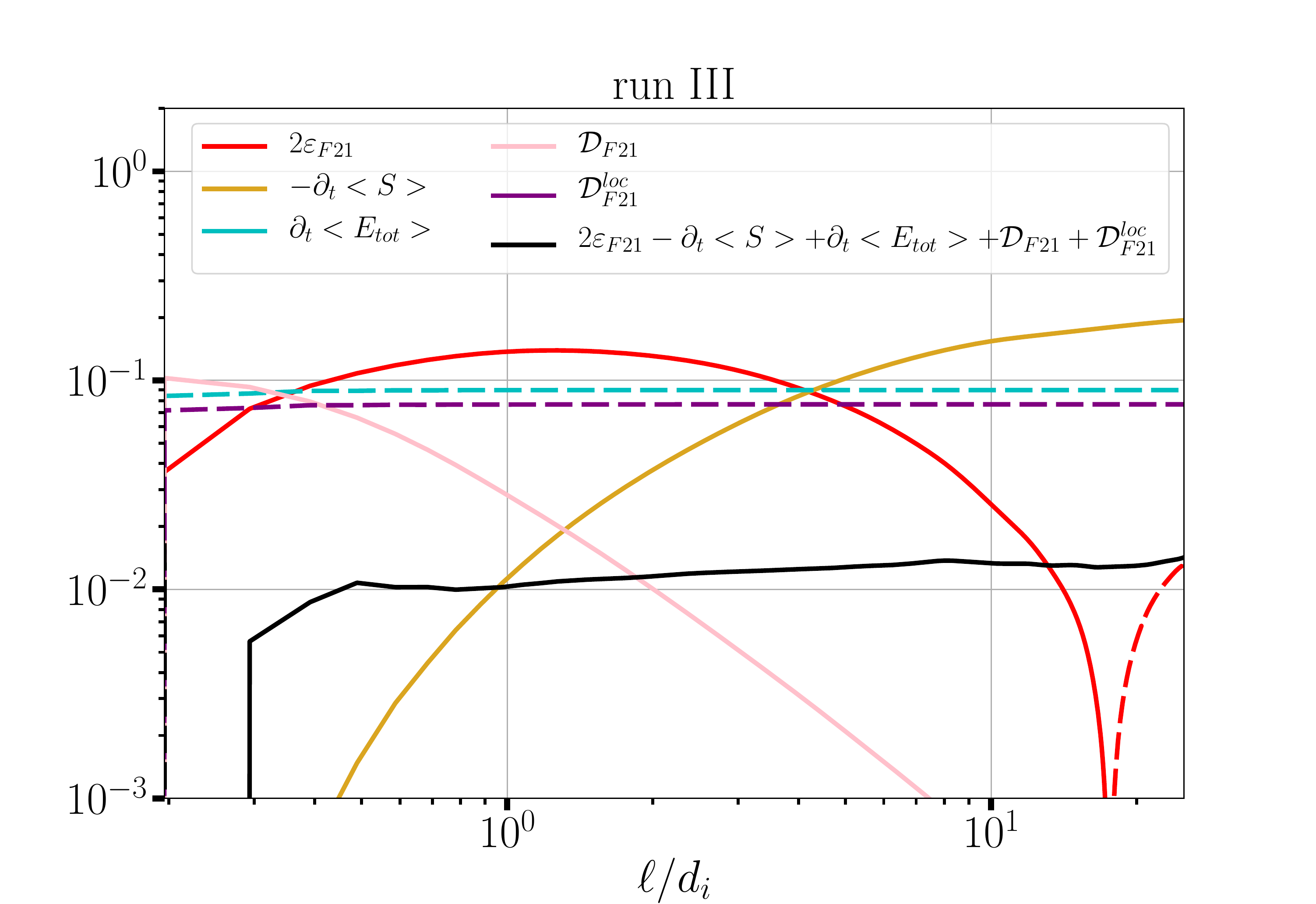}
	\caption{Calculation of equation (\ref{fullF21_comp2}) (black line) and its components for all three Runs. Plain lines represent positive values whereas dashed lines represent negative ones.}
	\label{runs_F21}
\end{figure}

Fig. \ref{runs_F21} shows that, again, the results of equation (\ref{fullF21_comp2}) do not change much between the three Runs. The local terms (i.e., independent on the increment $\el$) $\dFm$ and $\partial_t \lang E_{tot} \rang$ are almost equal, which can be intuited by the fact that $\dFm \sim -\varepsilon_{F21}$ in the limit of $\rho \sim \rho'$. Note that in this case the model still perfectly holds for Run I, whereas with equation (\ref{fullA18_comp2}) the term $ \partial_t \lang R_E + R_E' \rang$ showed an irregular increase at large scale. This may suggest that the model depicted by equation (\ref{fullF21_comp2}) is more robust than the previous one.

Interpreting the balance between the terms of the non-stationary F21 law is more subtle than for A18 because of the presence of the local terms $\partial_t \lang E^{tot} \rang$ and $\dFm$. Here it is the scale dependent term $-\partial_t \lang S \rang$ that acts as a ``forcing'' to input energy into the system, but this energy is diminished by the local (i.e., increment $\el=0$) energy variation term $\partial_t \lang E^{tot} \rang$ and the local dissipation $\dFm$. Then, the resulting energy balance is split, similarly to the A18 model, between the cascading part to small scales and the dissipation (both being scale dependent quantities):
\begin{equation} \label{fullF21_ir}
2\eF+\Big[-\partial_t \lang S \rang+\partial_t \lang E^{tot} \rang + \dFm \Big]+\dF =0.
\end{equation}

\section{Conclusion} \label{conclusion}

Similarly to the case of IHMHD, several exact laws exist for estimating the turbulence energy cascade rate in CHMHD. In this paper, making use of high resolution free-decay simulations of CHMHD turbulence, we showed that the two exact laws available for this model provide the same value of $\varepsilon$, as was already proven for IHMHD laws \citep{ferrand19}. In the absence of a direct mathematical proof of the equivalence between the two compressible laws, this paper brings evidence that they indeed describe the same turbulent cascade. The influence of the Hall effect on the energy cascade was also investigated. It appears that the development of a Hall-driven energy cascade in numerical simulations may be much more hindered by the action of dissipation at near-ion scales than by the size of the sub-ionic range. This underlines the potential importance of using a certain amount of hyperviscosity when doing simulations of CHMHD as a model of the large- and intermediate-scales in a plasma.

The question as to how the exact laws behave in absence of an external forcing led us to investigate a more general form of compressible exact laws, dropping the usual assumption of time stationarity and considering otherwise neglected time derivatives. This study shows that a shift in the interpretation of both laws occurs: instead of the continuous (in time) large scale forcing, the laws point toward the existence of a scale dependent reservoir of energy, mainly described by the time derivative terms, from which either the turbulent cascade or the dissipation (depending on the considered scale) draw. This reservoir overall coincides with the energy dissipation rate, which suggests that the considered cascading energy is ultimately bound to be fully dissipated.

These results confirm the non-trivial assertion that the final, well-known form of exact laws, obtained with the assumption of stationarity, remain valid within the inertial range even for decaying turbulent flows in which the aforementioned assumption is not verified, as it is the case in some turbulent space plasmas such as the SW taken far away from the sun.


\section*{Acknowledgements} 

This work was granted access to the HPC resources of CINES under allocation 2021 A0090407714 made by GENCI. N.A. acknowledge financial support from CNRS/CONICET Laboratoire International Associé (LIA) MAGNETO. N.A. acknowledges financial support from the following grants: PICT 2018 1095 and UBACyT 20020190200035BA.

\newpage
\clearpage


\section*{Appendix \\ axi-symmetric law calculation}

Throughout this paper, all terms of the generalized exact laws are computed using the isotropic decomposition described in section \ref{methods}. However, for simulations with $B_0 \neq 0$ the isotropy assumption should not \textit{a priori} be used. Instead we thus propose an axi-symmetric scheme that may prove to be more suited for the analysis of such simulation data.

We assume a symmetry of revolution around $\bb_0$, which is here aligned with the \textbf{z}-axis. In this case we adopt cylindrical coordinates: the increment vector is defined as $\el = (\ell_{\perp},\phi,\ell_z)$ and the derivative operator becomes $\nab_{\el} \cdot \lang {\textbf F} \rang = \frac{1}{\ell_{\perp}}\partial_{\ell_{\perp}} [\ell_{\perp} \lang F_{\ell_{\perp}} \rang(\ell_{\perp},\ell_z)] + \partial_{\ell_z}\lang F_{\ell_z}\rang (\ell_{\perp},\ell_z)$. However, the discrete decomposition adopted in this paper makes it impossible to effectively compute $\partial_{\ell_z}\lang F_{\ell_z}\rang (\ell_{\perp},\ell_z)$ at arbitrary values of $(\ell_{\perp},\ell_z)$ without resorting to multi-dimensional interpolation on irregular grids, bringing lots of additional calculations and more imprecision to the result. Consequently, we only consider the perpendicular component of the flux that is averaged over the parallel increments ${\ell_z}$, thus the derivative operator : $\nab_{\el} \cdot \lang {\textbf F} \rang = \frac{1}{\ell_{\perp}}\partial_{\ell_{\perp}} [\ell_{\perp} \lang F_{\ell_{\perp}} \rang(\ell_{\perp})]$. To do so we first compute the projection of $\textbf{F}$ on the direction of $\el_{\perp}$:
\begin{equation}
	\lang F_{\ell_{\perp}} \rang (\ell_{\perp},\phi,\ell_z) = \lang cos(\phi)F_x + sin(\phi)F_y\rang,
\end{equation}
and then take the average over all directions (i.e., over $\ell_z$ and $\phi$):
\begin{equation}
	\lang F_{\ell_{\perp}} \rang (\ell_{\perp}) = \sum_{\phi,\ell_z} \frac{\lang F_{\ell_{\perp}} \rang (\ell_{\perp},\phi,\ell_z)}{57}.
\end{equation}
Here we only use 57 directions, corresponding to all directions of the isotropic decomposition forming an angle of $45^\circ$ or more with the parallel direction. This is done to obtain results pertaining to all values of $\ell_\perp$ while avoiding redundant calculations due to the periodicity of the data cubes.

\begin{figure}
	\centering
	\includegraphics[width=\hsize,trim={40 35 70 40},clip]{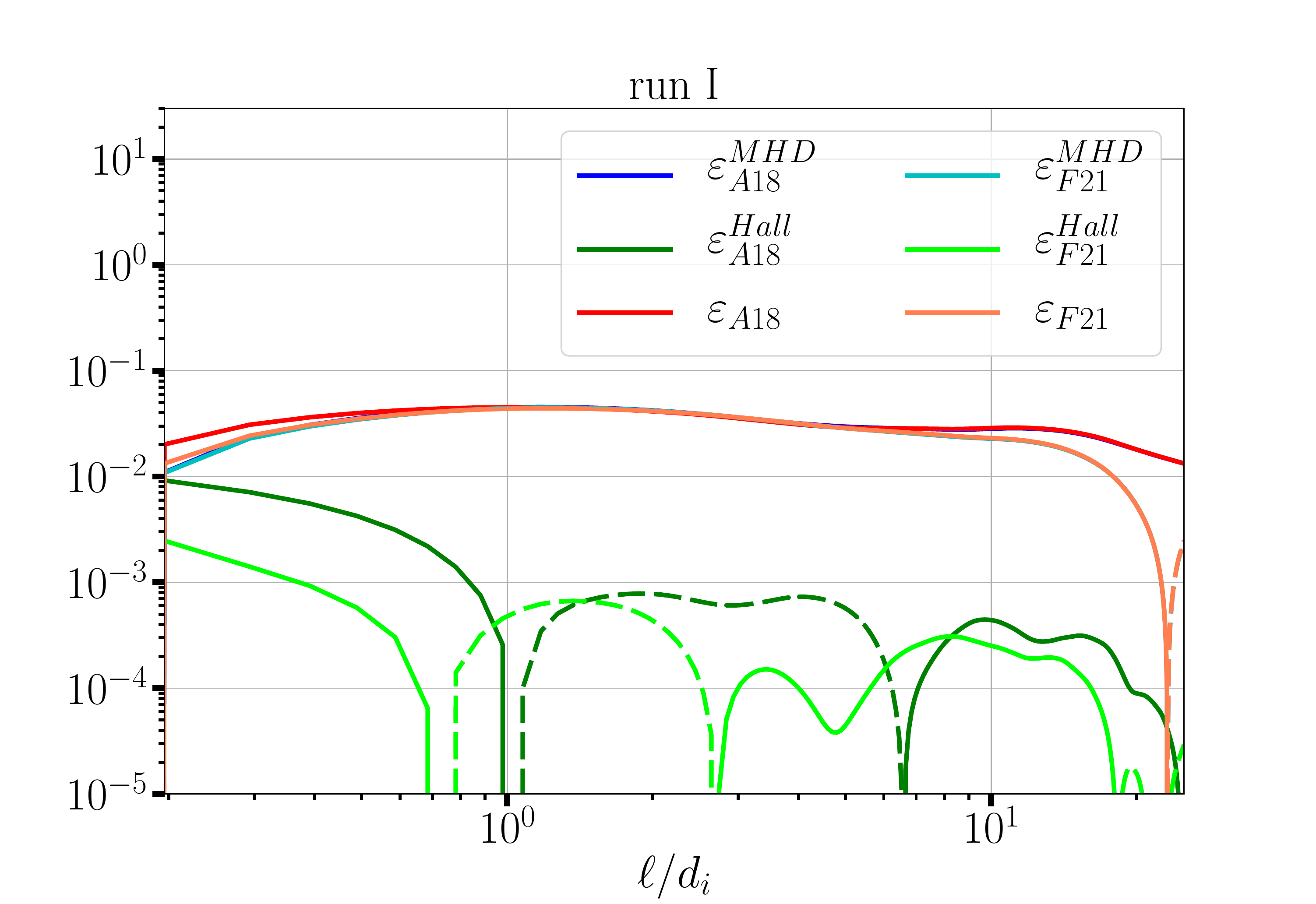}
	\caption{Energy cascade rate calculated with laws A18 and F21 for Run I, using the axi-symmetric method.}
	\label{run_I_axisym}
\end{figure}

This method presents two inconveniences: first, the statistics are weaker than the isotropic decomposition as we probe a smaller number of directions for the increment vector. This may lead to less precise calculations. Second, disregarding parallel fluxes may lead to miss a small portion of the cascade. Due to these limitations the isotropic decomposition may sometimes yield slightly better results even in presence of a background magnetic field. A good way to test the efficiency of this method is to compute the energy cascade rate of both exact laws A18 and F21 and see if they match as one would expect. This test was used on Run I and the results are reported in Fig. \ref{run_I_axisym}. We immediately observe that the two laws, and especially the Hall components, do not match as well as they are with the isotropic decomposition or for Run II (see Fig. \ref{runs_comp}), which may be a consequence of the aforementioned limitations. For this reason, we finally chose to keep using the isotropic decomposition to study Run I in this paper.

Still, note that Run I remains a weakly anisotropic simulation, with a background magnetic field of only 2. Thus, the conclusion drawn in this appendix may not be true for simulations featuring a stronger background field, i.e. stronger anisotropies. Ultimately, checking which method is the most suited one to lead the study should be done on a case-by-case basis.

\bibliography{Ref}
\end{document}